\documentclass[12pt,a4paper]{article}
\pdfoutput=1

\usepackage[utf8]{inputenc}
\usepackage[T1]{fontenc}

\usepackage{multirow}
\usepackage{cite}
\usepackage{amsmath}
\usepackage{color}
\usepackage{amsfonts}
\usepackage{amssymb}
\usepackage{graphicx}
\usepackage{geometry}
\usepackage{amssymb,epsfig,subfigure}
\usepackage{hyperref}
\usepackage{url}
\usepackage{comment}
\usepackage[font=footnotesize]{caption}
\usepackage{slashed}

\makeatletter
\renewcommand\section{\@startsection {section}{1}{\z@}%
                                 {-3.5ex \@plus -1ex \@minus -.2ex}
                                   {2.3ex \@plus.2ex}%
                                   {\normalfont\large\bfseries}}
\renewcommand\subsection{\@startsection{subsection}{2}{\z@}%
                                   {-3.25ex\@plus -1ex \@minus -.2ex}%
                                     {1.5ex \@plus .2ex}%
                                     {\normalfont\bfseries}}
\renewcommand\subsubsection{\@startsection{subsubsection}{3}{\z@}%
                                   {-3.25ex\@plus -1ex \@minus -.2ex}%
                                     {1.5ex \@plus .2ex}%
                                     {\normalfont\itshape}}
\makeatother

\def\pplogo{\vbox{\kern-\headheight\kern -29pt
\halign{##&##\hfil\cr&{\ppnumber}\cr\rule{0pt}{2.5ex}&\ppdate\cr}}}
\makeatletter
\def\ps@firstpage{\ps@empty \def\@oddhead{\hss\pplogo}%
  \let\@evenhead\@oddhead 
}
\def\maketitle{\par
 \begingroup
 \def\thefootnote{\fnsymbol{footnote}}
 \def\@makefnmark{\hbox{$^{\@thefnmark}$\hss}}
 \if@twocolumn
 \twocolumn[\@maketitle]
 \else \newpage
 \global\@topnum\z@ \@maketitle \fi\thispagestyle{firstpage}\@thanks
 \endgroup
 \setcounter{footnote}{0}
 \let\maketitle\relax
 \let\@maketitle\relax
 \gdef\@thanks{}\gdef\@author{}\gdef\@title{}\let\thanks\relax}
\makeatother

\numberwithin{equation}{section}

\newcommand\eea{\end{eqnarray}}
\newcommand\bea{\begin{eqnarray}}

\def\beq{\begin{equation}}
\def\eeq{\end{equation}}

\newcommand{\be}{\begin{equation}}
\newcommand{\ee}{\end{equation}}
\newcommand{\ba}{\begin{align}}
\newcommand{\ea}{\end{align}}
\newcommand{\bg}{\begin{gather}}
\newcommand{\eg}{\end{gather}}
\newcommand{\bseq}{\begin{subequations}}
\newcommand{\eseq}{\end{subequations}}

\renewcommand{\t}{\tilde}

\newcommand{\mc}{\mathcal}

\newcommand{\coment}[1]{}

\newcommand{\der}[2]{\frac{d#1}{d#2}}

\begin{document}
\setcounter{page}0
\def\ppnumber{\vbox{\baselineskip14pt
}}
\def\ppdate{
} \date{}

\author{Lucas Daguerre$^1$, Matias Ginzburg$^2$, Gonzalo Torroba$^{3, 2}$\\
[7mm] \\
{\normalsize \it $^1$ Center for Quantum Mathematics and Physics (QMAP)}\\
{\normalsize \it Department of Physics \& Astronomy, University of California, Davis, CA 95616 USA}\\
{\normalsize \it $^2$ Instituto Balseiro, UNCuyo and CNEA}\\
{\normalsize \it S.C. de Bariloche, R\'io Negro, R8402AGP, Argentina}\\
{\normalsize \it $^3$Centro At\'omico Bariloche and CONICET}\\
{\normalsize \it S.C. de Bariloche, R\'io Negro, R8402AGP, Argentina}\\
}

\bigskip
\title{\bf  Holographic entanglement entropy inequalities beyond strong subadditivity
\vskip 0.5cm}
\maketitle




\begin{abstract}
The vacuum entanglement entropy in quantum field theory provides nonperturbative information about renormalization group flows. Most studies so far have focused on the universal terms, related to the Weyl anomaly in even space-time dimensions, and the sphere free energy $F$ in odd dimensions. In this work we study the entanglement entropy on a sphere of radius $R$ in a large radius limit, for field theories with gravity duals. At large radius the entropy admits a geometric expansion in powers of $R$; the leading term is the well-known area term, and we also consider the subleading contributions.
These terms can be physical, they contain information about the full renormalization group flow, and they reproduce known monotonicity theorems in particular cases. We set up an efficient method for calculating them using the Hamilton-Jacobi equation for the holographic entanglement entropy. We first reproduce the known result for the area term, the coefficient multiplying $R^{d-2}$ in the entanglement entropy. We then obtain the holographic result for the $R^{d-4}$ term and establish its irreversibility.
Finally, we derive the $R^{d-6}$ coefficient for holographic theories, and also establish its irreversibility. This result goes beyond what has been proved in quantum field theory based on strong subadditivity, and hints towards new methods for analyzing the monotonicity of the renormalization group in space-time dimensions bigger than four.
\end{abstract}
\bigskip

\newpage

\tableofcontents

\vskip 1cm

\section{Introduction}\label{sec:intro}

Irreversibility theorems provide key insights into the nonperturbative structure of quantum field theories (QFTs). The first such theorem was proved by \cite{Zamolodchikov:1986gt}, and established the decrease of the central charge $C$ in $d=1+1$ dimensions along renormalization group flows. This theorem was rederived using quantum information tools in \cite{Casini:2006es}, and an extension of these methods allowed to establish the F-theorem in $d=2+1$ dimensions \cite{Casini_2012-F}. In $d=3+1$ dimensions, the A-theorem was proved by \cite{Komargodski:2011vj} using unitarity and the dilaton, and then in \cite{Casini:2017vbe} based on strong subadditivity of the entanglement entropy.

This situation is somewhat puzzling. On the one hand, these proofs use quite different methods (euclidean, Lorentzian and information theoretic), and it is not clear how they are related. One would hope for some unifying understanding, but it has not emerged so far. Furthermore, none of the results extend to more than 3+1 space-time dimensions. A clue that these issues should have a positive resolution comes from QFTs with holographic duals. Indeed, for such theories, the null energy condition (NEC) allows to construct a running C-function and establish the irreversibility of holographic renormalization group (RG) flows \cite{Freedman:1999gp, Myers:2010xs, Myers:2010tj}. The proposed C-function depends on the metric scale factor, and coincides with the universal A or F terms at fixed points. Further progress for understanding holographic RG flows in terms of the EE was made in \cite{Liu:2012eea,Liu:2013una}; we will revisit their method in the Appendix.

Motivated by these questions, in the present work we will analyze the entanglement entropy for holographic RG flows with the goal of finding new inequalities. We use holography because we want to access $d>4$ dimensions, and because we hope that this may provide clues for future field theoretic approaches. We will go beyond previous irreversibility results by obtaining new inequalities for the geometric large radius terms of the holographic entanglement entropy. As we will discuss, these contain physical information about the renormalization group and about the effective gravitational action induced by integrating out the QFT degrees of freedom.

We consider Poincare invariant QFTs in $d$ space-time dimensions. We assume that at short distances the theory is described by a conformal field theory; relevant perturbations trigger an RG flow, which we assume ends at an IR CFT (different from the UV one).  We will analyze the EE associated to a spherical region of radius $R$. Near a fixed point, it admits an expansion
\be\label{eq:Sfp}
S(R) = \mu_{d-2} R^{d-2}+ \mu_{d-4} R^{d-4}+ \ldots + \left \lbrace 
\begin{matrix}
(-1)^{\frac{d}{2}-1} 4A\,\log (R/\epsilon)\;,\;d \;\text{even}\\
(-1)^{\frac{d-1}{2}} F\;,\;d \;\text{odd}
\end{matrix}
\right.
\ee
$A$ and $F$ are the so-called universal terms (related to the Weyl anomaly in even space-time dimensions, and the sphere free energy in odd dimensions), and $\epsilon$ is a short distance cutoff. The $R^{d-2k}$ terms arise from geometric quantities of the boundary of the entangling region \cite{Casini:2018kzx}, and they will be the main focus in our work.

The entropy and its $\mu$ coefficients diverge as the short distance cutoff $\epsilon \to 0$. However, we will be interested in comparing two different theories: the UV fixed point one, and the theory that undergoes the nontrivial RG flow. We will review below how this leads to a finite, cutoff-independent, entropy differece $\Delta S$.

For an RG flow with a typical mass scale $m$, the UV CFT corresponds to  $m= 0$. In this limit, the sphere EE has the form (\ref{eq:Sfp}) with coefficients that we call $\mu_{d-2k}^{UV\,CFT}$. For $m R \gg 1$, the IR fixed point is approached, and the EE is also of the form (\ref{eq:Sfp}) but with different coefficients $\mu_{d-2k}^{IR}$.\footnote{Being an IR fixed point, the theory contains irrelevant deformations, and the typical mass scale $m$ acts as a UV cutoff for the IR effective theory.} We will be interested in the entropy difference
\be
\Delta S(R) =S^{IR}(R)-S^{UV\,CFT}(R)=  \Delta \mu_{d-2} R^{d-2}+\Delta \mu_{d-4} R^{d-4}+ \ldots\,,
\ee
with
\be
\Delta \mu_{d-2k}=\mu_{d-2k}^{IR}-\mu_{d-2k}^{UV\,CFT}\,.
\ee
As discussed in Sec. \ref{subsec:largeR}, this becomes independent of the cutoff and hence is a property of the continuum theory. The entropy difference appears naturally in information-theoretic approaches to irreversibility \cite{Casini:2016fgb, Casini:2016udt, Casini:2017vbe, Lashkari:2017rcl}.

The $\Delta A$ or $\Delta F$ are intrinsic to the fixed points, namely they are independent of the RG flow that connected the UV and IR CFTs. This is what ``universal'' means in this context. The other quantities $\Delta \mu_{d-2k}$ in the large $m R$ expansion do depend on the RG trajectory, but can still be physical (i.e. finite in the continuum limit). The simplest example is $\Delta \mu_{d-2}$, the coefficient of the area term. It corresponds to the renormalization of $1/G_N$ when weakly coupling gravity to the QFT; equivalently, the flow of the area term in the EE coincides with the low energy QFT contribution to black hole entropy. Ref. \cite{Casini:2014yca} showed that
\be
\Delta \mu_{d-2}= - \frac{\pi}{d(d-1)(d-2)} \int d^dx\,x^2 \langle \Theta(x) \Theta(0)
\rangle \,,\ee
where $\Theta(x)$ is the trace of the energy-momentum tensor. From here and reflection positivity or unitary, it follows that
\be
\Delta \mu_{d-2}\leq 0\;,\;d \ge 2\,.
\ee

This result for $\Delta \mu_{d-2}$ is also interesting because it contains the C-theorem as a particular case. Indeed, in the limit $d \to 2$, $\Delta \mu_{d-2} R^{d-2 } \to -\frac{1}{6}(c_{UV}- c_{IR})\,\log(m R)$, and
\be\label{eq:sumrule1}
c_{UV}- c_{IR}=3\pi \int d^2x\,x^2 \langle \Theta(x) \Theta(0)\rangle \geq 0\,.
\ee
Eq. (\ref{eq:sumrule1}), established first in \cite{Cappelli:1990yc}, is known as a sum-rule: the integrand in the right hand side depends on the RG trajectory, but its integral should only depend on the end-point central charges.

Further progress on understanding the coefficients in the expansion of the EE was made in \cite{Casini:2017vbe}, which showed that
\be
\Delta \mu_{d-4}\geq 0\;,\;d \ge 4\,.
\ee
The A-theorem is obtained for $d \to 4$. Given these results, it is natural to conjecture that the higher order terms satisfy
\be\label{eq:conjecture1}
(-1)^k\,\Delta \mu_{d-2k}\geq 0\,.
\ee
This would also imply the validity of A-theorems for all even $d$ by setting $2k=d$. So far there is no proof of this conjecture.

Our goal in this work is to analyze the $\Delta \mu_{d-2k}$ in theories with gravity duals. With this aim, we consider asymptotically AdS geometries that represent holographic RG flows. We will find that the Hamilton-Jacobi equation for the holographic EE as a function of the radius $R$ and a radial cutoff $\epsilon$ provides an efficient method for obtaining the $\mu_{d-2k}$.
The null energy condition will then allow us to establish new inequalities of the form (\ref{eq:conjecture1}). 
We will obtain holographic expressions for $\mu_{d-4}$ and $\mu_{d-6}$, and will prove (\ref{eq:conjecture1}) for $k=1, 2, 3$. The result for $k=3$ is also conceptually important as it goes beyond current results based on strong subadditivity. It also gives further support for the conjecture (\ref{eq:conjecture1}) which, however, we will not be able to prove in general.

The work is structured as follows.
In Sec. \ref{sec:setup} we describe the holographic setup. In Sec. \ref{sec:HJ} we derive the Hamilton-Jacobi (HJ) equation for the holographic EE, we discuss properties of the large radius expansion, and we compare this with the linearized holographic RG. Secs. \ref{sec:d2d4} and \ref{sec:d6} contain the main results -- the expressions for $\mu_{d-2}, \mu_{d-4}, \mu_{d-6}$ and the associated inequalities. Finally, in Sec. \ref{sec:concl} we present our conclusions and future directions. Additional explicit calculations that supplement the main text are given in the Appendices, where in particular we explore an alternative method based on solving the minimal surface equation and matching UV and IR approximations.

\section{Holographic setup}\label{sec:setup}

In this work we consider QFTs that are described by a UV CFT at very short distances, and a different IR CFT at long distances. The two fixed points are connected by an RG flow. For concreteness, we can think that this RG flow is triggered by perturbing the UV CFT by a single relevant deformation,
\be\label{eq:relevant}
S= S_\text{CFT,UV} + \int d^dx\,g\,\mc O(x)\,,
\ee
where $\mc O$ is a primary operator of dimension $\Delta <d$. The relevant coupling $g$ defines a mass scale typical for the RG, $m \sim g^{1/(d-\Delta)}$. Multiple relevant operators can also be added, and this will not change our analysis.
We will probe the properties of the RG flow using the entanglement entropy for a spherical region of radius $R$. We will focus on QFTs that admit gravity duals, where the EE is calculated by the Ryu-Takayanagi (RT) formula \cite{Ryu:2006bv, Ryu:2006ef, Lewkowycz:2013nqa}. Let us briefly describe this next.

\subsection{Gravity dual}

\begin{figure}[ht]
    \centering
    \includegraphics[width=0.7\textwidth]{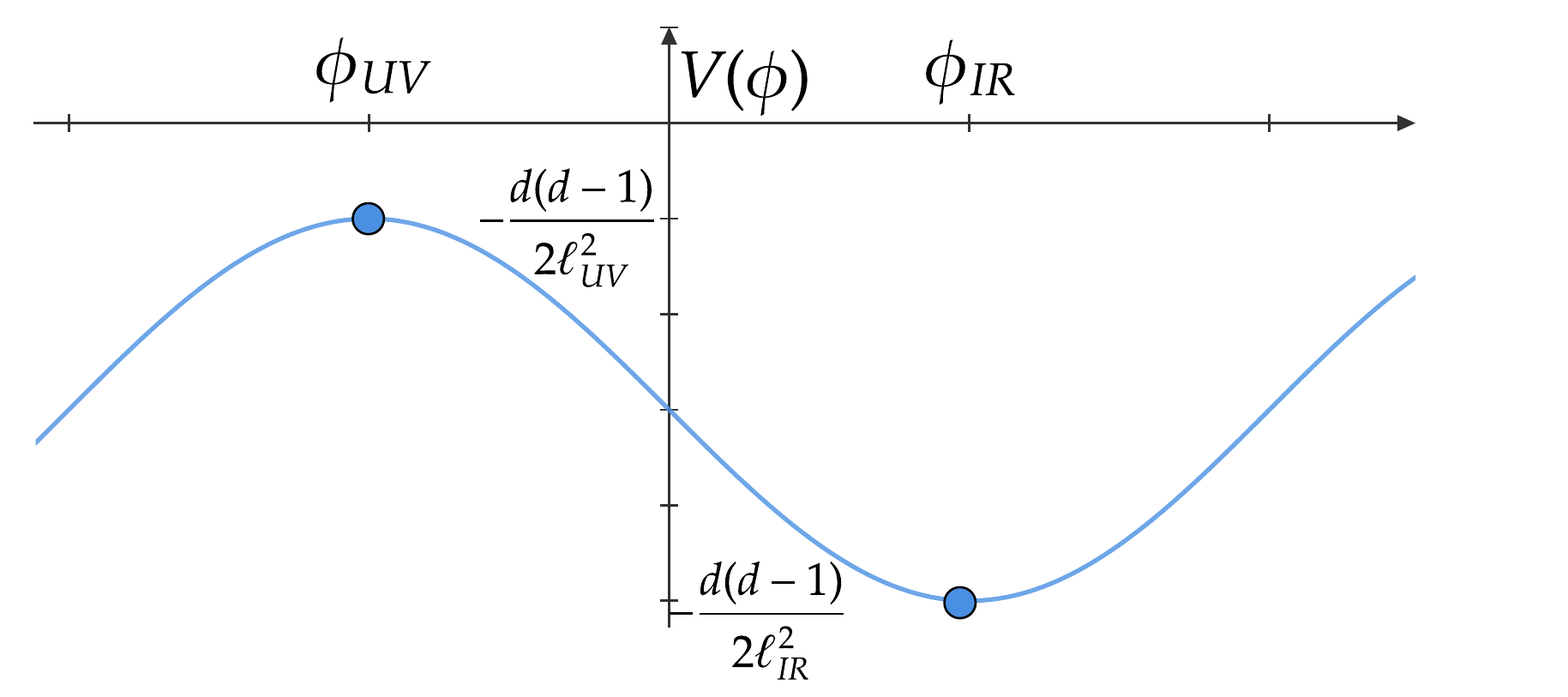}
    \captionsetup{width=0.9\textwidth}
    \caption{Schematic representation of the bulk potential that interpolates between the $\text{UV}$ and $\text{IR}$ fixed points for an arbitrary holographic RG flow. The conformal fixed points at $\phi_c$ have $V(\phi_c)=-\frac{d(d-1)}{2\ell_c^2}$ and are described by $\text{AdS}_{d+1}$ geometries with radius $\ell_c$ ($c=\text{UV},\text{IR}$).}
    \label{fig:HolographicPotential}
\end{figure}

The holographic dual is taken as Einstein-Hilbert gravity coupled to a scalar field $\phi$ dual to the relevant operator $\mc O$:
\be
S=\frac{1}{16\pi G_N^{(d+1)}} \int d^{d+1}x\,\sqrt{-g}\left( R^{(d+1)}- g^{MN} \partial_M \phi \partial_N \phi-V(\phi) \right)-\frac{1}{8\pi G_N^{(d+1)}}\int d^d x\; \sqrt{\gamma} K\,.
\ee
Additional scalars are straightforward to include.
In order to represent an arbitrary RG flow between two fixed points, the potential should have two critical points at $\phi_{c}$ such that $V(\phi_c)=-\frac{d(d-1)}{2\ell_c^2}$, where $c=\text{UV},\text{IR}$ (see Figure \ref{fig:HolographicPotential}). If $\phi=\phi_c$, the gravity solution is $AdS_{d+1}$ with radius $\ell_c$.
We require that the matter sector satisfies the NEC. This energy condition has been used previously in the holographic proof of the irreversibility of the universal term in arbitrary dimensions \cite{Freedman:1999gp, Myers:2010xs, Myers:2010tj}. Moreover it is the weakest of the most common local conditions \cite{Wall:2012uf}.


The solution to the scalar wave equation in AdS has two decay modes near the asymptotic boundary: a source-type term, and the VEV-type term \cite{Aharony:1999ti}. The relevant deformation (\ref{eq:relevant}) is dual to turning on a source in the asymptotic boundary. This perturbs $\phi$ away from $\phi_{UV}$, which will then evolve radially towards $\phi_{IR}$. The background that describes this (preserving Poincare symmetry in the boundary) can be parametrized as
\be \label{DomainWall}
    ds^2 =\frac{\ell_{UV}^2}{z^2} \left(\frac{dz^2}{f(z)}-(dx^0)^2+d\rho^2+\rho^2 d\Omega_{d-2}^2 \right)\,, \hfill \:\:\:\:\:\:\:\phi = \phi(z).
\ee

To proceed, we will only need the following properties. First, note that when $f(z)=1$ for all $z$, this geometry recovers the $\text{AdS}_{d+1}$ spacetime with radius $\ell_{\text{UV}}$. For arbitrary $f(z)$ we only require that $f(z)\to 1$ near the asymptotic boundary $z \to 0$. On the other hand, at large $z$, $f \to (\ell_{UV}/\ell_{IR})^2$. Also, the NEC for the matter sector translates on the equations of motions to the monotonicity requirement $f'(z)>0$. So in summary, the main properties of the scale factor are
\begin{equation}
\label{eq:NEC_conseq}
    1 \leq \; f(z)\; \leq \left(\frac{\ell_{\text{UV}}}{\ell_{\text{IR}}}\right)^2,\:\:\:\:\:\:\:\:
f'(z) \ge 0\,,
\end{equation}
with $f'(z)=0$ for $\text{AdS}_{d+1}$ spacetimes. See Appendix \ref{app:eoms} for more details.

\subsection{Holographic Entanglement Entropy}

For an entangling region $\Sigma$ in a constant time slice,  the holographic EE in asymptotically AdS spacetimes is given by the Ryu-Takayanagi prescription \cite{Ryu:2006bv,Ryu:2006ef}
\begin{equation}\label{RT}
    S=\text{min} \:\frac{\text{Area}(\sigma)}{4G_N^{(d+1)}},
\end{equation}
where $\sigma$ is a codimension-2 hypersurface in the bulk, anchored on $\partial \Sigma$ and  which minimizes the area; $G_N^{(d+1)}$ is the $(d+1)$-dimensional Newton constant. This result was established in \cite{Lewkowycz:2013nqa}.

\begin{figure}[ht]
    \centering
    \includegraphics[width=.3\textwidth]{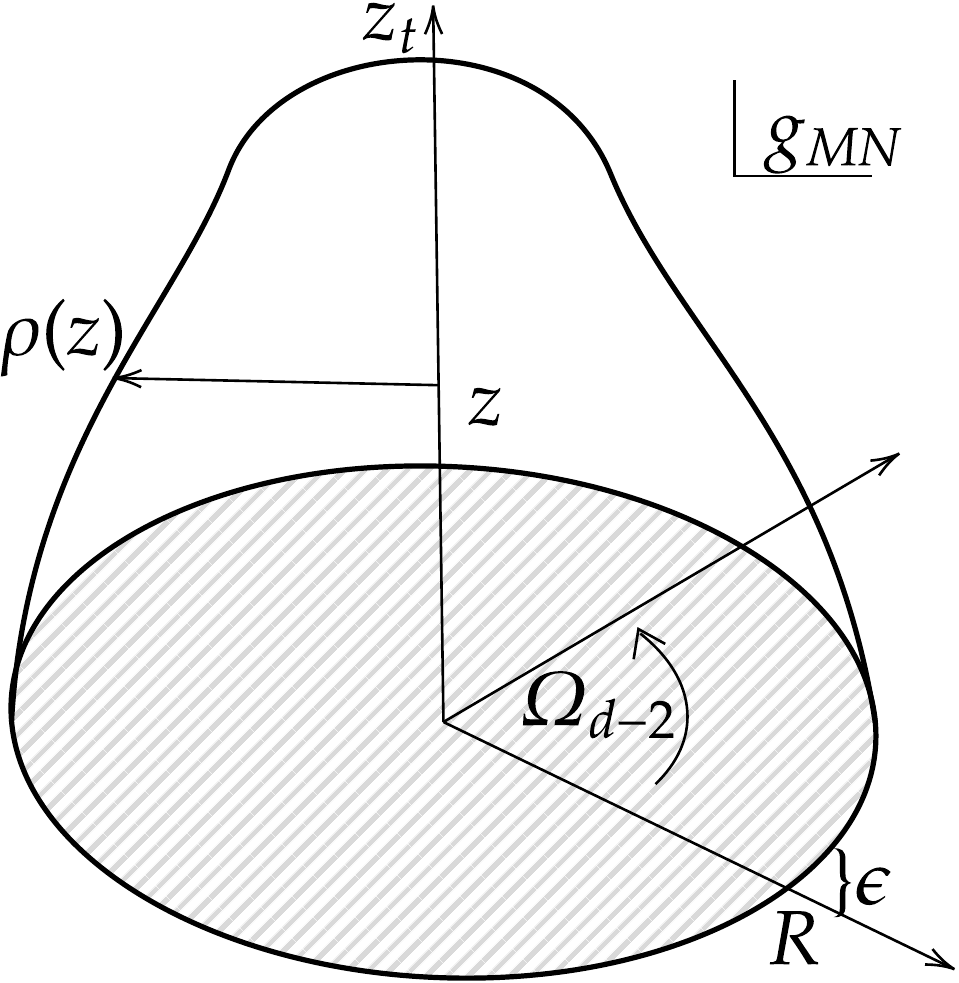}
    \captionsetup{width=0.9\textwidth}
    \caption{Representation  of the minimal surface $\sigma$ for a spherical region $\Sigma$ of radius $R$ anchored at $z=\epsilon$.}
    \label{fig:RT}
\end{figure}

To calculate the area of the minimal surface $\sigma$ we exploit the fact that the bulk metric \eqref{DomainWall} is diagonal and that the boundary is time independent and spherically symmetric. The minimal surface is a $(d-1)$-hypersurface parametrized by $d-2$ angular variables $\Omega_{i}$ and a radial function $\rho(z)$.  See Figure \ref{fig:RT} for a schematic representation.
Thus, $S$ is given by
\begin{equation} \label{eq:Sdef}
    S=\gamma_d \int_{\epsilon}^{z_t} dz\,\frac{\rho(z)^{d-2}}{z^{d-1}}\,\sqrt{\big( \rho'(z)\big)^2 + \frac{1}{f(z)}}\;,\;\;\;\gamma_d\equiv \frac{\text{Vol}(S^{d-2})\ell_{\text{UV}}^{d-1}}{4G_N^{(d+1)}}=\frac{2\pi^{\frac{d-1}{2}}}{\Gamma\left(\frac{d-1}{2}\right)}\frac{\ell_{\text{UV}}^{d-1}}{4G_N^{(d+1)}}\,,
\end{equation}
where $\epsilon$ is a geometric cutoff that regulates the entropy, and $z_t$ is a bulk radial turning point.\footnote{The radial cutoff $\epsilon$, which preserves Poincare invariance, is dual to the $TT$ deformation in the UV CFT \cite{McGough:2016lol}.} The turning point $z_t$ arises at $\rho=0$ as a consequence of spherical symmetry. The profile described by $\rho(z)$ has the
boundary condition
\be\label{eq:UVbc}
\rho(\epsilon)=R\,.
\ee
The turning point obeys
\be\label{eq:zt}
\rho(z_t)=0\,,\qquad \rho'(z_t)=-\infty\,.
\ee
Note also that when the size of the entangling region $R \to 0$, the minimal surface collapses to a point, and this gives zero area,
\be\label{eq:Sbc1}
\lim_{R \to 0^+} S =0\,.
\ee

 We must evaluate $S$ on the profile $\rho(z)$ that minimizes the area. This is the solution for $\rho(z)$ given by the Euler-Lagrange equation for \eqref{eq:Sdef}, 
\begin{equation}\label{eq:ELeq}
    \rho(z) z^{d-1} \sqrt{\rho'(z)^2+\frac{1}{f(z)}}\partial_z\left(\frac{\rho'(z)}{z^{d-1}\sqrt{\rho'(z)^2+\frac{1}{f(z)}}}\right) - \frac{(d-2)}{f(z)}=0.
\end{equation} 
This is a  non-linear second order ODE in $z$ for $\rho(z)$. It cannot be solved in closed form for a general metric $f(z)$ except for $d=2$. 

At large $R$, (\ref{eq:ELeq}) can be solved approximately in the UV and IR regions, and then both expressions can be matched in some overlapping regime \cite{Liu:2012eea, Liu:2013una}. We will discuss this in Appendix \ref{sec:Lagrange} . But we will find it more convenient to apply the Hamilton-Jacobi method at large $R$, and we turn to this next.

\section{Hamilton-Jacobi approach}
\label{sec:HJ}

The Hamilton-Jacobi method for a mechanical degree of freedom $q(t)$ gives an equation for its on-shell action $S$ as a function of initial data $q(t_0)=q_0$,
\be\label{eq:HJgen}
-\frac{\partial S(q_0, t_0)}{\partial t_0} = H \left(q_0, p_0 = \frac{\partial S}{\partial q_0}, t_0\right)\,,
\ee
with $H$ the Hamiltonian.
It appears naturally in holography with the radius playing the role of time, and the on-shell gravity action giving the large $N$ partition function of the dual field theory. In this context, the HJ formulation  is very convenient as it gives directly an equation for the on-shell action in terms of boundary data, without having to go through solving the bulk equations of motion.  It has been applied extensively to the holographic RG, starting from \cite{deBoer:1999tgo}, to Wilson loops \cite{Pontello:2015yla} and to the entanglement entropy \cite{Jackson:2013eqa, Jankowski:2020bmw}, among others.

\subsection{Derivation}\label{subsec:HJder}

We begin by briefly reviewing how to obtain (\ref{eq:HJgen}) for the holographic entropy (\ref{eq:Sdef}), explaining some subtle points specific to our case.

The solution to the equation of motion (\ref{eq:ELeq}) gives a function $\rho(z; \epsilon, R)$ that depends explicitly on the boundary data $\epsilon$ and $R$. Similarly, the on-shell entropy is a function $S=S(\epsilon, R)$. In general, the derivatives $\partial_z \rho$ and $\partial_\epsilon \rho$ are different. However, at $z=\epsilon$, we have $\rho(z=\epsilon; \epsilon, R)=R$. Since $R$ is independent of $\epsilon$, the total derivative $d\rho/d\epsilon=0$ at $z=\epsilon$, and hence
\be\label{eq:2der}
\frac{\partial \rho}{\partial \epsilon}(z;\epsilon, R) \Big |_{z=\epsilon}=- \frac{\partial \rho}{\partial z}(z;\epsilon, R) \Big |_{z=\epsilon}\,.
\ee
This relation is needed for the following derivation.

To proceed, we compute the variation $\partial_\epsilon S(\epsilon, R)$. This varies the endpoints as well as the integrand in (\ref{eq:Sdef}), which on-shell depends on $\epsilon$. There is no contribution from varying the endpoint $z_t$, because the integrand vanishes at $z=z_t$. On the other hand, varying the integrand, imposing the equation of motion and integrating by parts, gives a purely boundary term. Using here (\ref{eq:2der}) and combining with the contribution from the endpoint at $z=\epsilon$ obtains
\be\label{eq:partialS1}
\frac{\partial S}{\partial \epsilon}=- \frac{\gamma_d}{f(z)} \frac{\rho(z)^{d-2}}{z^{d-1}} \frac{1}{\sqrt{\big( \partial_z \rho\big)^2 + \frac{1}{f(z)}}} \Bigg|_{z=\epsilon}\,.
\ee
We recognize the right hand side as minus the Hamiltonian $H= L - \frac{\partial L}{\partial \rho'} \rho'$ at $z=\epsilon$.

To complete the derivation, we need the relation between the momentum conjugate to $\rho(z)$ and $\partial_R S$. A calculation very similar to the one just described gives
\be\label{eq:dSdR}
\frac{\partial S}{\partial R}=- \gamma_d \frac{\rho(z)^{d-2}}{z^{d-1}} \frac{\partial_z \rho}{\sqrt{\big( \partial_z \rho\big)^2 + \frac{1}{f(z)}}} \Bigg|_{z=\epsilon}\,.
\ee
Using this equation to write $\partial_z \rho$ in terms of $\partial_R S$ and replacing into (\ref{eq:partialS1}), we arrive at
\be\label{eq:HJ1}
\frac{\partial S}{\partial \epsilon}=-  \frac{1}{f(\epsilon)^{1/2}}\,\sqrt{\left(\gamma_d \frac{R^{d-2}}{\epsilon^{d-1}} \right)^2- \left(\frac{\partial S}{\partial R} \right)^2}\,.
\ee
This is the desired HJ equation for the holographic entanglement entropy. It has to be solved in the domain $\epsilon \in (0, \infty)$, $R \in (0, \infty)$ and, recalling (\ref{eq:Sbc1}), the boundary condition is
\be \label{eq>bcR0}
S(\epsilon, R \to 0^+)=0\,.
\ee
We will use it to solve directly for $S=S(\epsilon, R)$ without recourse to the equations of motion. 

A conceptually important point is that, while originally $\epsilon$ was introduced as a very small and fixed geometric cutoff, in the HJ equation $\epsilon$ is varied in $0<\epsilon <\infty$. This is very similar to what happens with the running Wilsonian cutoff, and is at the basis of the connection between the HJ equation and the holographic RG \cite{deBoer:1999tgo}.

We will also use an equivalent form of (\ref{eq:HJ1}),
\be\label{eq:HJ2}
f(\epsilon) \left(\frac{\partial S}{\partial \epsilon} \right)^2+\left(\frac{\partial S}{\partial R} \right)^2=\left(\gamma_d \frac{R^{d-2}}{\epsilon^{d-1}} \right)^2\,.
\ee
This is an eikonal equation, and it would be interesting to explore potential connections with the bit thread formulation \cite{Freedman:2016zud}, something that we leave for future work.\footnote{A similar result for the gravitational action is interpreted as the WKB limit of the bulk Wheeler-de Witt equation \cite{deBoer:1999tgo, Heemskerk:2010hk}.}

\subsection{Exact solution for pure AdS}\label{subsec:exact}

It is in general not possible to solve explicitly the HJ equation (\ref{eq:HJ1}) for arbitrary $f(z)$, but we will shortly see that it can be easily solved in a large $R$ expansion. Before turning to that, however, we will focus on the case of pure $\text{AdS}_{d+1}$, $f(z)=1$ for all $z$, where an exact solution can be obtained.

We solve the equation by proposing that the entropy depends on the dimensionless combination $y(\epsilon, R)=\sqrt{\left(\frac{R}{\epsilon}\right)^{2}+1}$, $S(\epsilon, R)=S(y(\epsilon, R))$.\footnote{This particular combination is actually $z_t/\epsilon$ and is motivated by the solution using the Euler-Lagrange equation in Appendix \ref{sec:Lagrange}, where it is clear that the surface that minimizes the area is a cut of a sphere of radius $\sqrt{R^2+\epsilon^2}$.} For $R=0$ there is no entanglement entropy in this new variable so the boundary condition \eqref{eq>bcR0} is written as 
\begin{equation}\label{bcAds}
    S_{AdS}(y=1)=0.
\end{equation}
Now the new equation in one variable is integrable
\begin{equation}
    \frac{d S_{AdS}}{d y}=\gamma_{d}\left(y^{2}-1\right)^{\frac{d-3}{2}}\,.
\end{equation}
Using the condition \eqref{bcAds} we get the final result
\begin{equation}
S_{AdS}(y)=\gamma_{d} \int_{1}^{y} d \tilde{y}\left(\tilde{y}^{2}-1\right)^{\frac{d-3}{2}}.
\end{equation}
The result is the incomplete beta function\footnote{The incomplete beta function $\displaystyle \text{B}(x;a,b)=\int_0^x dt\:t^{a-1}(1-t)^{b-1}$ for $a,b>0$ and $0 \leq x \leq 1$.}
\begin{equation} \label{eq:S_UV}
\begin{split}
    S_{AdS}(y) =& \gamma_d  \int_{1/y}^{1}  d\omega\:\:\:\omega^{1-d}(1-\omega^2)^{\frac{d-3}{2}} \\
    =& \frac{\gamma_d}{2}\left(-\text{B}\left(\frac{1}{1+\left(\frac{R}{\epsilon}\right)^2};\frac{2-d}{2},\frac{d-1}{2}\right)+\frac{\Gamma(\frac{2-d}{2})\Gamma(\frac{d-1}{2})}{\sqrt{\pi}}\right).
\end{split}
\end{equation}

The expansion in powers of $\frac{\epsilon}{R}$ gives 
\begin{equation}
\begin{split}
-\frac{1}{2}\text{B}\left(\frac{1}{1+\left(\frac{R}{\epsilon}\right)^2};\frac{2-d}{2},\frac{d-1}{2}\right)&=\left(\frac{R}{\epsilon}\right)^{d-2}\Bigg(\frac{1}{(d-2)}-\frac{1}{2(d-4)}\left(\frac{\epsilon}{R}\right)^2+\ldots\\
&+(-1)^{k+1}\frac{(2k-3)!!}{(2k-2)!!}\frac{1}{(d-2k)}\left(\frac{\epsilon}{R}\right)^{2(k-1)}+\ldots\Bigg).
\end{split}
\end{equation}
We recognize the structure (\ref{eq:Sfp}), with
\begin{equation}\label{eq:muAdS}
\mu^{UV\,CFT}_{d-2k}(\epsilon) =
    (-1)^{k+1}\frac{(2k-3)!!}{(2k-2)!!}\frac{1}{(d-2k)}\frac{1}{\epsilon^{d-2k}}\,.
\end{equation}
In particular, the logarithmic universal term is recovered for even $d=2k$ recalling that
\begin{equation}
\frac{\Gamma(\frac{2-d}{2})\Gamma(\frac{d-1}{2})}{2\sqrt{\pi}}=\left((-1)^{k}\frac{(2k-3)!!}{(2k-2)!!}\right)\frac{1}{(d-2k)}+\mathcal{O}\Big((d-2k)\Big),
\end{equation}
and
\begin{equation}
\lim_{d\to 2k}\:\frac{x^{(d-2k)}-1}{(d-2k)}=\log(x).
\end{equation}

It is worth noting that these coefficients are different from the ones obtained in the original calculation \cite{Ryu:2006ef} --except for the universal terms that are identical. This is due to a different radial cutoff: in \cite{Ryu:2006ef}, the integral starts at $z=\epsilon$, but the boundary condition is taken as $\rho(z=0)=R$. In our case, we instead impose $\rho(z=\epsilon)=R$. One could view $\epsilon$ simply as a regulator, always taking $\epsilon \to 0$, and then we have just two different regularization schemes. However, in the HJ approach it is important that we vary $\epsilon$ over all scales in order to solve the HJ equation. So it is not necessarily the smallest scale in the problem. In fact, the EE with the cutoff procedure of \cite{Ryu:2006ef} does not satisfy the HJ equation (\ref{eq:HJ2}). We instead prefer to view the introduction of the Dirichlet wall at $z=\epsilon$ as a physical deformation of the boundary theory, akin to the $TT$ deformation \cite{McGough:2016lol}.

\subsection{Large $R$ expansion}\label{subsec:largeR}

We will now use the HJ equation to solve for the entropy in the large $R$ expansion. Let us first analyze some properties of this expansion.

The RG flow introduces some characteristic mass scale $m \sim g^{1/(d-\Delta)}$ (see (\ref{eq:relevant})), which in the gravity side translates into the value of $z$ for which $f(z)$ starts to differ appreciably from the UV value $f(0) =1$. At long distances, the theory flows to an IR fixed point; in gravity language, $f(z \to \infty) \sim \ell_{UV}^2/\ell_{IR}^2$. This regime is probed by the EE with radius $R \gg 1/m$. As reviewed in Sec. \ref{sec:intro}, near a fixed point the EE admits a geometric expansion
\bea\label{eq:SlargeR2}
S(\epsilon, R) &=&\gamma_d \left( \mu_{d-2}(\epsilon) R^{d-2}+ \mu_{d-4}(\epsilon) R^{d-4}+ \mu_{d-6}(\epsilon) R^{d-6}+ \ldots\right) \nonumber\\
&=&\gamma_d \left( \sum_{k=1}^{\lfloor d/2 \rfloor} R^{d-2k}\mu_{d-2k}(\epsilon) + \ldots\right)\,.
\eea
In the semiclassical bulk description, the leading contribution to the EE is proportional to $\gamma_d$, as in (\ref{eq:Sdef}).\footnote{In dual QFT variables, $\gamma_d \sim C_T$, the coefficient in the stress tensor two-point function. For instance, in $\mathcal N=4$ $SU(N)$ super Yang-Mills, $\gamma_d \sim N^2$.} To simplify the following formulas, we have made $\gamma_d$ explicit here in front of the large $R$ expansion. The coefficients $\mu_{d-2k}$ here then differ from their QFT counterparts (\ref{eq:Sfp}) by this factor of $\gamma_d$.

The expansion is valid for positive integer powers of $R$. The terms that decay with $R$ (the `$\ldots$' in the last line) generically have non-integer powers; these are related to the specific dimensions of the leading irrelevant operators that control the approach to the IR fixed point. We will not consider such terms in this work; Refs. \cite{Liu:2012eea,Liu:2013una} studied them in holographic theories.

The entropy is dimensionless, so $\mu_{d-2k}$ has dimensions of $1/(\text{length})^{d-2k}$ to compensate. On dimensional grounds then, for a dimensionless function $\hat{\mu}_{d-2k}$
\be
\mu_{d-2k} = \frac{1}{\epsilon^{d-2k}} \hat \mu_{d-2k}(g \epsilon^{d-\Delta})\,.
\ee
These expressions are not known in general; however, their parametric dependence may be understood near a fixed point using conformal perturbation theory. In particular, near the UV fixed point $\epsilon \ll g^{1/(d-\Delta)}$ we expect, up to order one constants,
\be
\mu_{d-2k} \sim \frac{1}{\epsilon^{d-2k}}+ \frac{1}{\epsilon^{d-2k}} (g^2 \epsilon^{2(d-\Delta)})+\ldots+ g^{\frac{d-2k}{d-\Delta}}\;,\;\text{for}\;\;\epsilon \ll g^{1/(d-\Delta)}\,,
\ee
where we used that the leading perturbation arises at order $g^2 \langle \mathcal O \mathcal O \rangle$ \cite{Rosenhaus:2014woa}.

The first term is an ultraviolet divergence coming from the UV CFT. The simplest example is the divergent area term, but such divergences also afflict other terms in the geometric large $R$ expansion. The second term, on the other hand, is a consequence of the relevant perturbation. Recalling that $\Delta<d$, this term will diverge for $\epsilon \to 0$ in the window
\be
\frac{d+2k}{2}< \Delta < d\,.
\ee
These divergences are familiar from renormalizable interactions in QFT. They may also be understood from the gravitational action induced by the QFT.
We will instead restrict to
\be\label{eq:window}
\Delta < \frac{d+2k}{2},
\ee
so that there are no UV divergences associated to the relevant perturbation. While the $\mu$'s are cutoff dependent, the difference $\Delta \mu_{d-2k}$ is finite as $\epsilon \to 0$, 
\be
\Delta \mu_{d-2k} \sim g^{\frac{d-2k}{d-\Delta}}\,.
\ee 
Therefore, by comparing the large radius expansion of the UV fixed point and of the theory with nontrivial RG flow we obtain physical (cut-off independent) entropy coefficients $\Delta \mu_{d-2k}$. They are a property of the continuum theory.\footnote{Instead of restricting $\Delta$ as in (\ref{eq:window}) there is another possibility. It was found in \cite{Casini:2018kzx} that the strong subadditivity formulas also work if one compares the EE $S(R)$ to the entropy of the CFT plus appropriate `counterterms' whose role is to cancel the UV divergences. However, this procedure in the gravity dual gives rise to expressions whose sign is not fully determined by the NEC. So we will not pursue this approach here.} The inequalities below will refer to these coefficients.

Let us now obtain the differential equations for the coefficients $\mu_{d-2k}$. For this, we replace (\ref{eq:SlargeR2}) into (\ref{eq:HJ1}) and set to zero the coefficient multiplying each independent power $R^{d-2k}$. Denoting $\epsilon \to z$ here, this gives linear differential equations $\mu'_{d-2k}(z)$ in terms of lower $\mu$'s. In particular, the first three are
\bea\label{eq:firsteqs}
 \partial_z \mu_{d-2}&=&- \frac{1}{z^{d-1} f(z)^{1/2}} \nonumber\\
\partial_z \mu_{d-4}&=& \frac{(d-2)^2}{2} \frac{z^{d-1}}{f(z)^{1/2}} \mu_{d-2}(z)^2 \\
 \partial_z \mu_{d-6}&=& \frac{(d-2)^4}{8} \frac{z^{3(d-1)}}{f(z)^{1/2}} \mu_{d-2}(z)^4+(d-2) (d-4) \frac{z^{d-1}}{f(z)^{1/2}}  \mu_{d-2}(z) \mu_{d-4}(z)\,. \nonumber
 \eea
Therefore, the HJ equation in the large $R$ expansion gives first order equations that can be solved iteratively. The area coefficient $\mu_{d-2}$ is sourced by the inhomogeneous area term in the HJ equation, and the higher order coefficients are in turn sourced by a nontrivial $\mu_{d-2}$. In Secs. \ref{sec:d2d4} and \ref{sec:d6} we will solve these equations and establish inequalities for the resulting solutions.

\subsection{Linearized approximation}\label{subsec:linear}

The Wilsonian RG can often be seen as an infinitesimal step in the exact RG \cite{Wilson:1973jj, Polchinski:1983gv, Rosten:2010vm} and holographic RG equations \cite{deBoer:1999tgo, Heemskerk:2010hk, Faulkner:2010jy}. It is then interesting to consider a linearized version of the EE HJ equation in order to derive a flow interpretation. The connection with the RG was stressed in \cite{Jackson:2013eqa}.

With this aim, let us consider the first step in the large radius expansion, writing
\be
S(\epsilon,R) =\gamma_d \left( S_{area}(\epsilon, R) + \hat S(\epsilon, R)\right)\,,
\ee
where $S_{area}$ cancels the inhomogenous right hand term in (\ref{eq:HJ2}),
\be
S_{area}(\epsilon,R) = R^{d-2} \int_\epsilon^\infty\,\frac{dz}{z^{d-1}f(z)^{1/2}}=R^{d-2} \mu_{d-2}(\epsilon)\,.
\ee
The HJ equation for $\hat S$ then reads
\be
-2 \frac{R^{d-2}}{\epsilon^{d-1}} f(\epsilon)^{1/2} \partial_\epsilon \hat S+ \frac{2(d-2)}{R} S_{area} \partial_R \hat S+ f(\epsilon) (\partial_\epsilon \hat S)^2+(\partial_R \hat S)^2 = - \frac{(d-2)^2}{R^2} S_{area}^2\,.
\ee
This is still the complete HJ equation now for $\hat S$; it contains both linear and quadratic terms, and shows that $S_{area}$ acts like a source term for $\hat S$.

The idea now is to neglect the terms that are quadratic in derivatives of $\hat S$,
\be\label{eq:linearized1}
-2 \frac{R^{d-2}}{\epsilon^{d-1}} f(\epsilon)^{1/2} \partial_\epsilon \hat S+ \frac{2(d-2)}{R} S_{area} \partial_R \hat S\approx  - \frac{(d-2)^2}{R^2} S_{area}^2\,.
\ee
We will check the validity of this approximation shortly. This equation may be interpreted as a geometric flow equation for the EE, relating the change in cutoff to the geometric deformation $\partial_R \hat S$ and the area term.

The linear differential equation can be solved by the method of characteristics. The general solution is
\bea
\hat S_{ren}(\epsilon,R)&=& - \frac{(d-2)^2}{2}  \int_\epsilon^\infty\,dv \,\frac{v^{d-1}\mu_{d-2}(v)^2}{f(v)^{1/2}}\left[R^2+2(d-2) \int_v^\epsilon du\,\frac{u^{d-1}}{f(u)^{1/2}} \mu_{d-2}(u)\right]^{\frac{d-4}{2}} \nonumber\\
&&
 + G\left(\sqrt{R^2+2(d-2) \int_{\epsilon_0}^\epsilon dv\,\frac{v^{d-1}}{f(v)^{1/2}} \mu_{d-2}(v)}\right)\,,
\eea
where $G(\eta)$ is some arbitrary function we should fix with boundary conditions, and $\epsilon_0$ comes from a choice of integration constant in the method.

In order to assess the validity of the linearized approximation, we expand $\hat S$ at large $R$ in terms of the $\mu_{d-2p}$, and plug into (\ref{eq:linearized1}). This gives ODEs term by term in the $1/R$ expansion; the first are
\bea\label{eq:lineareqs}
-2 z \sqrt{f(z)} \mu_{d-4}'(z)+(d-2)^2 z^d \mu_{d-2}(z)^2&=&0\nonumber\\
-2 z \sqrt{f(z)}\mu_{d-6}'(z)+2 (d-4) (d-2) z^d \mu_{d-2}(z) \mu_{d-4}(z)&=&0\nonumber\\
-2 z \sqrt{f(z)} \mu_{d-8}'(z)+2 (d-6) (d-2) z^d \mu_{d-2}(z) \mu_{d-6}(z)&=&0\,.
\eea
As opposed to the full nonlinear ODEs for the $\mu_{d-2p}$, these keep only terms that are linear in $(\mu_{d-4}, \mu_{d-6}, \ldots)$.

Relating (\ref{eq:firsteqs}) and (\ref{eq:lineareqs}), we find that $\mu_{d-4}$ is captured completely by the linearized approximation, while from the order $\mu_{d-6}$ the linearized approximation fails. The quadratic terms that are neglected (\ref{eq:linearized1}) turn out to be comparable to (\ref{eq:lineareqs}) beginning at $O(R^{d-6})$, and for this reason the linearized approximation is not useful for such non-universal terms. The equations (\ref{eq:firsteqs}) are still linear differential equations in the unknowns, and hence resemble RG beta functions for the entropy coefficients. But getting these right requires taking into account more nonlinear terms beyond the leading area contribution.

\subsection{Sphere free energy}\label{subsec:freenergy}

The large $R$ expansion (\ref{eq:SlargeR2}) must be modified for  odd space-time dimensions $d$ and has to include a term $F(\epsilon)$,
\begin{equation}
    S(\epsilon, R) =\gamma_d \left( \mu_{d-2}(\epsilon) R^{d-2}+ \mu_{d-4}(\epsilon) R^{d-4}+ \mu_{d-6}(\epsilon) R^{d-6}+ \ldots\right)+(-1)^{\frac{d-1}{2}}F(\epsilon).
\end{equation}
Inserting this new expansion in the Hamilton-Jacobi equation (\ref{eq:HJ2}) gives
\begin{equation}
    F'(\epsilon)=0,
\end{equation}
therefore $F$ is a constant. In a CFT this is the constant term of the free energy of a $d$ dimensional euclidean sphere \cite{Dowker:2010yj, Casini:2011kv}. 

The term $F$ can be fully determined at fixed points in our holographic context from the expression (\ref{eq:S_UV}) for the entropy  when $f(z)$ is constant. More precisely,
\begin{equation}
    S^{\text{UV}} \supset \gamma_d \frac{\Gamma(\frac{2-d}{2})\Gamma(\frac{d-1}{2})}{2\sqrt{\pi}},\:\:\:\:\:\:\:\:S^{\text{IR}} \supset \gamma_d \left(\frac{\ell_{\text{IR}}}{\ell_{\text{UV}}}\right)^{d-1} \frac{\Gamma(\frac{2-d}{2})\Gamma(\frac{d-1}{2})}{2\sqrt{\pi}},
\end{equation}
are the constant terms in the UV and IR expansions. This gives the inequality\footnote{The holographic value for $F$ matches with the universal term $a^*_d=\frac{\pi^{\frac{d}{2}}}{\Gamma\left(\frac{d}{2}\right)}\left(\frac{\ell}{\ell_P}\right)^{d-1}$ defined in \cite{Myers:2010xs, Myers:2010tj} via the identification $F=2\pi a^*_d$,  $\ell_P^{d-1}=8\pi G_N^{(d+1)}$ and recalling that $\frac{(-1)^{\frac{d-1}{2}}\Gamma\left(1-\frac{d}{2}\right)}{\pi}=\frac{1}{\Gamma\left(\frac{d}{2}\right)}$ due to the Euler's reflection formula.}
\begin{equation}
    \Delta F=F^{\text{IR}}-F^{\text{UV}}=\frac{(-1)^{\frac{d-1}{2}}\pi^{\frac{d-2}{2}}\Gamma\left(\frac{2-d}{2}\right)}{4G_N^{(d+1)}}\left(\ell_{\text{IR}}^{d-1}-\ell_{\text{UV}}^{d-1}\right)\leq 0 \:\:\:\:\:\:\:\:\: \forall d\:\text{odd},
\end{equation}
because of $\ell_{\text{UV}}\geq \ell_{\text{IR}}$ as a consequence of the NEC (\ref{eq:NEC_conseq}) and the positivity of the prefactor for $d$ odd. This F-theorem was obtained by \cite{Myers:2010xs, Myers:2010tj} in an holographic analysis for all $d$ odd and by \cite{Casini_2012-F,Casini:2017vbe} using QFT methods involving the SSA in $d=3$. At the moment there is no QFT proof for F-theorems in $d>3$ spacetime dimensions.

\section{Analysis of the $R^{d-2}$ and $R^{d-4}$ terms}\label{sec:d2d4}

In this section we calculate the $\mu_{d-2}(z)$ and $\mu_{d-4}(z)$ terms in the large $R$ expansion of the EE. The area term has been calculated for holographic theories in \cite{Casini:2015ffa} using a different method based on stress-tensor correlators; we will reproduce their result. The holographic result for $\mu_{d-4}(z)$ has not appeared in the literature, as far as we are aware.

For each coefficient we give expressions in terms of a single integral involving $f(z)$ that allow to prove the inequalities $\Delta \mu_{d-2}\leq 0$ and $\Delta \mu_{d-4} \geq 0$ for $\Delta \mu_{d-2k}=\mu_{d-2k}-\mu_{d-2k}^{UV\,CFT}$ using the NEC (\ref{eq:NEC_conseq}). 
We will also rewrite these coefficients in a way that makes manifest the anomaly result when $d\to 2$ and $d \to 4$ respectively. This ``anomaly oriented'' procedure will also be useful for our analysis of $\Delta \mu_{d-6} \leq 0$ in Sec.\ref{sec:d6}.

\subsection{$\Delta \mu_{d-2}(z)$}\label{subsec:mud2}

The equation for $\mu_{d-2}(z)$ is
\be
\mu'_{d-2}(z)=-\frac{1}{z^{d-1}f(z)^{1/2}}\,.
\ee
The integration constant should be independent of $R$, and $z$; moreover for $z=\epsilon \to 0$, this must reproduce the UV CFT expansion $\mu_{d-2}^{UV\,CFT}(\epsilon)\sim \epsilon^{-(d-2)}$ (\ref{eq:muAdS}). Therefore, this
integrates to
\begin{equation}
\label{eq:mu_d_2}
   \mu_{d-2}(z) = \int_z^\infty  \frac{dv}{v^{d-1} f(v)^{1/2}} > 0\,.
\end{equation}
Subtracting the contribution of pure AdS, $f(v)=1$, gives the inequality for the area term
\begin{equation}\label{eq:Dmu_d_2}
    \Delta \mu_{d-2}(z)=\int_z^\infty  \frac{dv}{v^{d-1} f(v)^{1/2}}-\int_z^\infty  \frac{dv}{v^{d-1} }\leq 0,
\end{equation}
because $f(z) \geq 1$ for all $z$. The decrease of the area term coefficient along RG flows has been proved in QFT in \cite{Casini:2014yca, Casini:2016udt}. The holographic calculation using stress-tensor correlators was carried out in \cite{Casini:2015ffa}. Our HJ method is extremely simple and gives the same result.\footnote{To match their result, the appropriate change of variables is $A(r)=\log\left(\frac{\ell_{\text{UV}}}{z}\right)$, $A'(r)=\frac{\sqrt{f(z)}}{\ell_{\text{UV}}}$.}

In the limit $d \to 2$, (\ref{eq:Dmu_d_2}) should reproduce the holographic C-theorem, and so it should only depend on UV and IR CFT quantities and not on the full RG flow. To exhibit this, we integrate by parts the $1/v^{d-1}$ factor,
\begin{equation}
\label{eq:mu_d_2_anom}
    \mu_{d-2}(z) = \frac{1}{d-2} \frac{1}{z^{d-2} f(z)^{1/2}}-\frac{1}{2}\frac{1}{d-2} g_{d-2}(z)\,,
\end{equation}
where we define
\begin{equation}
    g_{d-2}(z) \equiv \int_z^\infty \frac{dv}{v^{d-2}}\,\frac{f'(v)}{f(v)^{3/2}} \geq 0\,.
\end{equation}
For future use, note that since $\mu_{d-2}(z)>0$ for all $z$,
\begin{equation}
\label{eq:bound_g_d_2}
    0 \leq g_{d-2}(z) < \frac{2}{z^{d-2}f(z)^{1/2}}\,.
\end{equation}
Finally, taking the limit $z=\epsilon \to 0$ and recalling that $f(\epsilon) \to 1$ and $g_{d-2}=0$ for AdS,\footnote{In general $\Delta \mu_{d-2k}$ has potentially the most divergent contributions coming from terms that goe like $g_{d-2k}(z)=\int_z^\infty \frac{dv}{v^{d-2k}}\,\frac{f'(v)}{f(v)^{(2k+1)/2}}$ when $z\to 0$. However because $f(z)=1+(\mu z)^{2(d-\Delta)}+...$ in the UV limit (\ref{f(z)}), the integrand for such contributions goes like $\sim v^{d-2\Delta +2k-1}$. In the window $\Delta < \frac{d+2k}{2}$ (\ref{eq:window}) the integrand decays slower than $v^{-1}$ for $v\to 0 $, so $\Delta \mu_{d-2k}$ remains finite and cut-off independent when $\epsilon \to 0$ as discussed in Sec. \ref{subsec:largeR}.}
\begin{equation}
    \Delta \mu_{d-2}=-\frac{1}{2}\frac{1}{d-2}g_{d-2}(0)\leq 0\,.
    \end{equation}
When $d \to 2$, $g_{d-2}$ becomes a total derivative and the logarithmic term in $\Delta S$ is recovered,
\begin{equation}
  \frac{\Delta c}{3}=  R\frac{d \Delta S}{dR} \Big |_{R^{d-2}}=\gamma_2\left(f(\infty)^{-\frac{1}{2}}-f(0)^{-\frac{1}{2}}\right)=\frac{\ell_{\text{IR}}}{2G_N^{(3)}} -\frac{\ell_{\text{UV}}}{2G_N^{(3)}}  = \frac{c_{\text{IR}}-c_{\text{UV}}}{3}\leq 0,
\end{equation}
giving raise to the weak version of the holographic C-theorem. The central charge is identified via $c = \frac{3\ell}{2G_N^{(3)}}$  in $d=2$\cite{Brown:1986nw}. 
\subsection{$\Delta \mu_{d-4}(z)$}\label{subsec:mud4}

After the previous warm-up, we now move to the $R^{d-4}$ coefficient, for which there was no known holographic expression. Starting from the equation,
\begin{equation}\label{eq:eqmud4}
    \mu_{d-4}'(z)= \frac{(d-2)^2}{2} \frac{z^{d-1}}{f(z)^{1/2}} \mu_{d-2}(z)^2\,,
\end{equation}
the solution integrates to
\begin{equation}
\label{eq:mu_d_4}
     \mu_{d-4}(z) = -\frac{(d-2)^2}{2} \int_z^\infty dv\,\frac{v^{d-1}}{f(v)^{1/2}} \left( \int_v^\infty  \frac{du}{u^{d-1} f(u)^{1/2}} \right)^2 \,.
\end{equation}

The expression (\ref{eq:mu_d_4}) satisfies $\mu_{d-4}<0$ for all $z$. Therefore, the NEC gives the following inequality,
\begin{equation}\label{eq:Dmu_d_4}
\begin{split}
    \Delta \mu_{d-4}(z)&=-\frac{(d-2)^2}{2} \left[\int_z^\infty dv\,\frac{v^{d-1}}{f(v)^{1/2}} \left( \int_v^\infty  \frac{du}{u^{d-1} f(u)^{1/2}} \right)^2-\int_z^\infty dv\,v^{d-1} \left( \int_v^\infty  \frac{du}{u^{d-1}} \right)^2\right]\\
     & \geq 0    \,.
\end{split}   
\end{equation}
Indeed, the AdS contribution is explicitly larger than the one corresponding to the holographic RG flow, since $f(z)>1$ is a monotonically increasing function.

Ref. \cite{Casini:2017vbe} proved $\Delta \mu_{d-4} \le 0$ in QFT using the strong subadditivity for multiple boosted spheres. Here we have obtained the same result for theories with gravity duals and the NEC, in the large $R$ expansion. It would be interesting to relate (\ref{eq:mu_d_4}) to euclidean boundary correlators. This would help towards an euclidean proof of the A-theorem, with a corresponding sum rule.

Following what we did before for $\mu_{d-2}$, let us rewrite (\ref{eq:Dmu_d_4}) in order to make the holographic A-theorem manifest when $d \to 4$. From (\ref{eq:mu_d_2_anom}), we have
\be
\mu_{d-2}(z)^2=\frac{1}{(d-2)^2} \frac{1}{z^{2(d-2)}f(z)}-\frac{g_{d-2}(z)}{4(d-2)^2}\left(\frac{4}{z^{d-2}f(z)^{1/2}}-g_{d-2}(z)\right)\,.
\ee
Replacing into (\ref{eq:eqmud4}) and integrating,
\be
\mu_{d-4}(z)=-\frac{1}{2}\int_z^\infty  \frac{dv}{v^{d-3} f(v)^{3/2}}+\frac{1}{8}\int_z^{\infty}dv\frac{v^{d-1}}{f(v)^{1/2}}g_{d-2}(v)\left(\frac{4}{v^{d-2}f(v)^{1/2}}-g_{d-2}(v)\right)\,.
\ee
Finally, integrating by parts with respect to $1/v^{d-3}$ in the first term, and defining
\begin{equation}
g_{d-4}(z) \equiv \int_z^\infty \frac{dv}{v^{d-4}}\,\frac{f'(v)}{f(v)^{5/2}} \geq 0\,,
\end{equation}
we arrive at
\begin{equation}
\label{eq:mu_d_4_anom}
\begin{split}
    \mu_{d-4}(z)&=- \frac{1}{2(d-4)} \frac{1}{z^{d-4}f(z)^{3/2}}+ \frac{3}{4(d-4)} g_{d-4}(z)\\
    &\qquad \qquad+\frac{1}{8}\int_z^{\infty}dv\frac{v^{d-1}}{f(v)^{1/2}}g_{d-2}(v)\left(\frac{4}{v^{d-2}f(v)^{1/2}}-g_{d-2}(v)\right).
\end{split}
\end{equation}
Note that from the integration by parts,
\be
\frac{1}{2}\int_z^\infty  \frac{dv}{v^{d-3} f(v)^{3/2}}= \frac{1}{d-4} \frac{1}{z^{d-4}f(z)^{3/2}}- \frac{3}{2(d-4)} g_{d-4}(z) \ge 0
\ee
we obtain the bound
\begin{equation}
\label{eq:bound_g_d_4}
0 \leq g_{d-4}(z) < \frac{2}{3} \frac{1}{z^{d-4}f(z)^{3/2}}\,.
\end{equation}
Finally setting $z=\epsilon \to 0$ and subtracting the pure AdS result, we obtain the desired inequality
\be\label{eq:Dmud4}
    \Delta \mu_{d-4}=\frac{3}{4(d-4)} g_{d-4}(0)+\frac{1}{8}\int_0^{\infty}dv\frac{v^{d-1}}{f(v)^{1/2}}g_{d-2}(v)\left(\frac{4}{v^{d-2}f(v)^{1/2}}-g_{d-2}(v)\right) \ge 0\,.
\ee
The last term is positive semidefinite because of (\ref{eq:bound_g_d_2}). It contributes by a finite amount to the entropy and depends on the full RG flow. 
In the window (\ref{eq:window}) for $k=2$, this is finite for $\epsilon \to 0$ and hence we have a physical cut-off independent coefficient.

When $d \to 4$, $g_{d-4}$ becomes a total derivative and the logarithmic term in $\Delta S$ is recovered,
\begin{equation}
   -4 \Delta a= R \frac{d \Delta S}{dR}\Big |_{R^{d-4}}=\frac{\gamma_4}{2}\left(f(0)^{-\frac{3}{2}}-f(\infty)^{-\frac{3}{2}}\right)=-\frac{\pi \ell^3_{\text{IR}}}{2G_N^{(5)}}+\frac{\pi \ell^3_{\text{UV}}}{2G_N^{(5)}}=-4(a_{\text{IR}}-a_{\text{UV}})\geq 0,
\end{equation}
where the $a$-anomaly is given by $a=\frac{\pi \ell^3}{8G_N^{(5)}}$ in $d=4$. The extra terms in (\ref{eq:Dmud4}) contribute to the constant nonuniversal term when $d\neq 4$.

\section{Beyond the SSA: the $R^{d-6}$ term}
\label{sec:d6}

In this section we give an integral expression of $\mu_{d-6}(z)$ in terms of $f(z)$ and provide a proof of the inequality $\Delta \mu_{d-6}\leq 0$ following an anomaly oriented procedure. This is the first term in the large $R$ expansion whose RG flow is not constrained by the boosted strong subadditivity. Surpisingly, we find that the NEC in the large $R$ expansion is sufficient to establish the inequality.

For the present case the equation is
\begin{equation}\label{eq:diffeqmud6}
    \mu_{d-6}'(z)= \frac{(d-2)^4}{8} \frac{z^{3(d-1)}}{f(z)^{1/2}} \mu_{d-2}^4(z)+(d-2) (d-4) \frac{z^{d-1}}{f(z)^{1/2}}  \mu_{d-2}(z) \mu_{d-4}(z)\,.
\end{equation}
Unlike what happens with the coefficients in Sec. \ref{sec:d2d4}, this is the first time there is a competition between two terms of opposite sign (because of (\ref{eq:mu_d_2}) and (\ref{eq:mu_d_4})). Using (\ref{eq:mu_d_2_anom}) and (\ref{eq:mu_d_4_anom}), expanding the two terms in (\ref{eq:diffeqmud6}), and integrating,
\begin{equation}
    \begin{split}
       \mu_{d-6}(z)
        &=\int_z^{\infty}dv\:\bigg[-\frac{1}{8}\frac{1}{v^{d-5}f(v)^{1/2}}+\frac{1}{4}\frac{v^3}{f(v)^2}g_{d-2}(v)-\frac{1}{16}\frac{v^{d+1}}{f(v)^{3/2}}g_{d-2}(v)^2\\
        &\:\:\:\:\:\:\:\:\:\:\:\:\:\:\:\:\:\:\:\:\:\:\:\:\:\:\:\:\:\:\:\:\:\:\:\:\:\:-\frac{1}{128}\frac{v^{3(d-1)}}{f(v)^{1/2}}g_{d-2}^2(v)\left(\frac{4}{u^{d-2}f(u)^{1/2}}-g_{d-2}(u)\right)^2\bigg]\\
        &\:\:\:+\int_{z}^{\infty}dv\:\bigg[\frac{1}{2}\frac{1}{v^{d-5}f(v)^{1/2}}-\frac{1}{4}\frac{v^3}{f(v)^2}g_{d-2}(v)-\frac{1}{2}\frac{v^{d-1}}{f(v)^{1/2}}\left(\frac{2}{v^{d-2}f(v)^{1/2}}-g_{d-2}(v)\right)\\
        &\:\:\:\:\:\:\:\:\:\:\:\:\:\:\times \left(\frac{3}{4}g_{d-4}(v)+\frac{(d-4)}{8}\int_v^{\infty}du\frac{u^{d-1}}{f(u)^{1/2}}g_{d-2}(u)\left(\frac{4}{u^{d-2}f(u)^{1/2}}-g_{d-2}(u)\right)\right) \bigg]\,.
    \end{split}
\end{equation}
Adding up the first two terms arising from both integrals gives
\begin{equation}
\label{eq:int_part_mu_d_6}
\frac{3}{8}\int_{z}^{\infty} \frac{dv}{v^{d-5}f(v)^{1/2}}=\frac{3}{8(d-6)} \frac{1}{z^{d-6} f(z)^{5/2}}-\frac{15}{16(d-6)}g_{d-6}(z) >0\,,
\end{equation}
where
\begin{equation}
    g_{d-6}(z) \equiv \int_z^\infty \frac{dv}{v^{d-6}}\,\frac{f'(v)}{f(v)^{7/2}}\,,
\end{equation}
and because of (\ref{eq:int_part_mu_d_6})
\begin{equation}
\label{eq:bound_g_d_6}
    0 \leq g_{d-6}(z)< \frac{2}{5}\frac{1}{z^{d-6}f(z)^{5/2}}\,.
\end{equation}
Finally, we arrive at
\begin{equation}
\label{eq:mu_d_6_anom}
    \begin{split}
        \mu_{d-6}(z)&=\frac{3}{8(d-6)} \frac{1}{z^{d-6} f(z)^{5/2}}-\frac{15}{16(d-6)}g_{d-6}(z) -\int_z^{\infty}dv\:\Bigg[\frac{1}{16}\frac{v^{d+1}}{f(v)^{3/2}}g_{d-2}(v)^2\\
        &+\frac{1}{128}\frac{v^{3(d-1)}}{f(v)^{1/2}}g_{d-2}(v)^2\left(\frac{4}{u^{d-2}f(u)^{1/2}}-g_{d-2}(u)\right)^2+\frac{1}{2}\frac{v^{d-1}}{f(v)^{1/2}}\left(\frac{2}{v^{d-2}f(v)^{1/2}}-g_{d-2}(v)\right)\\
        &\:\:\:\:\:\:\:\:\:\:\:\:\:\:\times \left(\frac{3}{4}g_{d-4}(v)+\frac{(d-4)}{8}\int_v^{\infty}du\frac{u^{d-1}}{f(u)^{1/2}}g_{d-2}(u)\left(\frac{4}{u^{d-2}f(u)^{1/2}}-g_{d-2}(u)\right)\right) \Bigg]\,.
    \end{split}
\end{equation}

Upon setting $z=\epsilon \to 0$, and subtracting the pure AdS contribution, we find 
\begin{equation}\label{eq:Dmud6}
    \begin{split}
        \Delta \mu_{d-6}&=-\frac{15}{16(d-6)}g_{d-6}(0) -\int_{0}^{\infty}dv\:\Bigg[\frac{1}{16}\frac{v^{d+1}}{f(v)^{3/2}}g_{d-2}(v)^2+\frac{1}{128}\frac{v^{3(d-1)}}{f(v)^{1/2}}g_{d-2}^2(v)\\
        &\:\:\:\:\:\times \left(\frac{4}{u^{d-2}f(u)^{1/2}}-g_{d-2}(u)\right)^2+\frac{1}{2}\frac{v^{d-1}}{f(v)^{1/2}}\left(\frac{2}{v^{d-2}f(v)^{1/2}}-g_{d-2}(v)\right)\\
        &\:\:\:\:\:\:\:\:\:\:\:\times \left(\frac{3}{4}g_{d-4}(v)+\frac{(d-4)}{8}\int_v^{\infty}du\frac{u^{d-1}}{f(u)^{1/2}}g_{d-2}(u)\left(\frac{4}{u^{d-2}f(u)^{1/2}}-g_{d-2}(u)\right)\right) \Bigg]\,.
    \end{split}
\end{equation}
Given (\ref{eq:window}), this is finite when $\epsilon \to 0$.
From to the inequalities (\ref{eq:bound_g_d_2}) and (\ref{eq:bound_g_d_4}), it satisfies the inequality
\be\label{eq:ineqd6}
\Delta \mu_{d-6} \leq 0\,.
\ee
Therefore, we have found that the NEC implies the decrease of the $R^{d-6}$ term along holographic RG flows. Finally, 
when $d \to 6$ the function $g_{d-6}$ becomes a total derivative and the logarithmic term is recovered
\begin{equation}
     R \frac{d \Delta S}{dR}\Big |_{R^{d-6}}=\frac{3}{8}\gamma_6\left(f(0)^{-\frac{5}{2}}-f(\infty)^{-\frac{5}{2}}\right)=\frac{\pi^2 \ell^5_{\text{IR}}}{4G_N^{(7)}}-\frac{\pi^2 \ell^5_{\text{UV}}}{4G_N^{(7)}}\leq 0.
\end{equation}

It would be interesting to translate (\ref{eq:Dmud6}) into stress tensor correlators. This expression already has the desired features of satisfying the inequality (\ref{eq:ineqd6}) and reproducing a sum rule for $d \to 6$. So it may provide hints towards an A-theorem for $d=6$.

\section{Conclusions and future directions}\label{sec:concl}

In this work we analyzed the coefficients $\Delta \mu_{d-2k}$ in the large radius expansion (\ref{eq:Sfp}) of the EE for field theories with gravity duals. These coefficients are finite for RG flows triggered by operators with dimension $\Delta < \frac{d+2}{2}$. Unlike the universal $A$ and $F$ terms, the $\Delta \mu_{d-2k}$ contain information about the full RG flow connecting the UV and IR fixed points. Starting from the Ryu-Takayanagi formula, we derived a Hamilton-Jacobi equation (\ref{eq:HJ2}) for the holographic EE $S(\epsilon, R)$. The cut off $\epsilon$ modifies the EE not just by cutting the integral, but also modifying the minimal surfaces, allowing $\epsilon$ to take any value. At large radius, the HJ equation reduces to first order differential equations for the $\mu_{d-2k}$, which are straightforward to solve. We derived explicit holographic results for $\mu_{d-2}, \mu_{d-4}, \mu_{d-6}$ and established the inequalities
\be
\Delta \mu_{d-2}  \le 0\;,\;\Delta \mu_{d-4}  \ge 0\;,\;\Delta \mu_{d-6}  \le 0\,.
\ee
The first two inequalities have been proved before for general QFTs using the boosted strong subadditivity \cite{Casini:2017vbe} (SSA). On the other hand, the inequality $\Delta \mu_{d-6} \le 0$ is new and goes beyond the SSA.

Let us end by discussing  future directions opened by these results. We have obtained integral expressions for $\Delta \mu_{d-4}$ and $\Delta \mu_{d-6}$ as a function of the metric scale factor. These expressions are a starting point for deriving expressions in terms of euclidean stress-tensor correlators. This would be interesting for different reasons. It could suggest a sum rule for the 4d A-theorem, generalizing the sum rule of $d=2$ (which uses $\langle T^\mu_\mu(x) T^\nu_\nu(0) \rangle$) to $d=4$. The $d=6$ result in terms of stress-tensor correlators, on the other hand, may shed light on a possible A-theorem for that dimensionality.

Another important lesson is that the NEC together with a large radius expansion give rise to the inequality $\Delta \mu_{d-6} \le 0$ that is stronger than current results using SSA. It would be very useful to translate this into quantum information conditions. It suggests thinking in terms of a large radius expansion in field theory.

Finally, it would be important to push the holographic analysis to the higher coefficients $\Delta \mu_{d-2k}$. Our results support the conjecture
\be
(-1)^k\,\Delta \mu_{d-2k}\geq 0\,,
\ee
and it would be nice to prove this for theories with gravity duals. The analysis for $\Delta \mu_{d-8}$ and higher coefficients appears to be significantly more involved, possibly requiring new tools so that it can be made systematic.


\section*{Acknowledgments}
We thank H. Casini, M. Huerta, R. Trinchero and M. Rangamani for comments on our work, and especially H. Casini for extensive discussions. We also thank H. Casini and M. Rangamani for comments on the final manuscript. LD is supported by a Dean's Distinguished Graduate Fellowship from the College of Letters and Science of the University of California, Davis. MG is supported by CNEA and UNCuyo, Inst. Balseiro. GT is supported by CONICET (PIP grant 11220200101008CO), ANPCyT (PICT 2018-2517), CNEA, and UNCuyo, Inst. Balseiro.

\appendix

\section{Gravity dual equations of motion}
\label{app:eoms}

The action for an arbitrary minimally coupled scalar field $\phi(z)$ in the bulk is
\begin{equation}
    S=\frac{1}{16\pi G_N^{(d+1)}} \int d^d x\;dz\;\sqrt{-g} \left(R^{(d+1)}- g^{MN} \partial_M \phi \partial_N \phi -V(\phi) \right) - \frac{1}{8\pi G_N^{(d+1)}}\int d^d x\; \sqrt{\gamma} K.
\end{equation}
Here we consider the Einstein-Hilbert action with the standard Gibbons-Hawking boundary term and some arbitrary potential $V(\phi)$ for the scalar field. The equations of motion are
\begin{equation}\label{EinstainScalarEOM}
    R_{MN}-\frac{1}{2} g_{MN} R = \partial_M \phi \partial_N \phi - \frac{1}{2} g_{MN} \left[(\partial _z\phi)^2+V(\phi) \right],
\end{equation}
and
\begin{equation}\label{scalarEOM}
    \frac{1}{\sqrt{-g}} \partial_M \left(\sqrt{-g} g^{MN}\partial_N \phi\right)-\frac{1}{2} \der{V}{\phi} =0.
\end{equation}

If the potential has a critical point $\der{V}{\phi}(\phi_c)=0$ with $V(\phi_c)=-\frac{d(d-1)}{2\ell_c^2}$, then pure $\text{AdS}_{d+1}$ with radius $\ell_c$ is a solution in the class of (\ref{DomainWall}) for the equations of motion with
\begin{equation}
    f(z)=\left(\frac{\ell_{\text{UV}}}{\ell_c}\right)^2  \; ,\;\:\: \phi(z)=\phi_c\:\:\:\:\:\: \forall z.
\end{equation}

Replacing the metric \eqref{DomainWall} in the equations of motion \eqref{EinstainScalarEOM} and \eqref{scalarEOM}\footnote{For the metric (\ref{DomainWall}) we find
$
R_{zz} = \frac{d}{2} \frac{(-2 f+ z f')}{z^2 f}$ and $R_{\mu\nu} = \eta_{\mu\nu} \frac{(-2d f+zf')}{2z^2}
$.} and combining them yields two coupled nonlinear ODEs in $z$
\begin{equation}\label{EinsteinEqScalar}
\phi''(z)- \left( \frac{d-1}{z}-\frac{f'(z)}{2f(z)}\right)\phi'(z)- \frac{\ell_{\text{UV}}^2}{2z^2 f(z)} \der{V(\phi)}{\phi} = 0,
\end{equation}
\begin{equation}\label{EinsteinEqlog}
\frac{d-1}{z} \frac{f'(z)}{f(z)} = 2 \phi'(z)^2.
\end{equation}

These equations cannot be solved in closed form for a general potential in arbitrary $d$, but notable features emerge after expanding the solutions near the critical points. Around the $\text{UV}$ fixed point the bulk potential can be expanded as
\begin{equation}
    V(\phi)\approx -\frac{d(d-1)}{2\ell_{\text{UV}}^2} + m^2 \phi^2 + \ldots,
\end{equation}
so that the equations of motion can be solved order by order. To begin with, at zeroth order $f(z)=1$ for all $z$ and 
\begin{equation} \label{boundary-phi}
    \phi(z) \approx \phi_{\text{UV}} z^{d-\Delta} + \phi_\Delta z^\Delta\:\:\:\:\:\:\:\:\:\;z\to 0\:\:,\;\:\:\:\:\:m^2\ell_{\text{UV}}^2 =\Delta (\Delta-d),
\end{equation}
solves (\ref{EinsteinEqScalar}). The constant $\phi_\Delta$ corresponds to the normalizable fall-off and it is proportional to the vacuum expectation value (VEV) of the field which is set to zero; $\phi_{\text{UV}}$ is dual to the source of the relevant boundary operator $\mathcal{O}$ of conformal dimension $\Delta$ and so the mass is tachyonic, \textit{viz.} $m^2 = \Delta(\Delta-d)<0$. This explains why  the $\text{UV}$ is a local maximum in Figure \ref{fig:HolographicPotential}. Subsequently equation \eqref{EinsteinEqlog} can be integrated using  $f(0)=1$,
\begin{equation}
     f(z) \approx 1 + \frac{d-\Delta}{d-1} \phi_{\text{UV}}^2  z^{2(d-\Delta)}.
\end{equation}

A similar procedure can be followed around the $\text{IR}$ fixed point. The limit $z\to\infty$ corresponds to the $\text{IR}$ asymptotic region where $\phi(z\to\infty)=\phi_{IR}z^{d-\t\Delta}$. The irrelevant boundary operator $\mathcal{O}$ has conformal dimension $\tilde{\Delta}$ with $d-\tilde{\Delta}<0$. The zeroth order in $f(z)$ is given by $f(\infty)= \ell_{UV}^2/\ell_{IR}^2$. This analysis justifies the following generic expansion for $f(z)$
\begin{equation} \label{f(z)}
    f(z) \approx \begin{cases}
    1 + (\mu z )^{2\alpha} +\ldots& z\to 0\:\:\:\:\: (\text{UV}) \\
    \frac{\ell_{UV}^2}{\ell_{IR}^2} \left( 1 - \frac{1}{(\tilde{\mu} z)^{2\tilde{\alpha}}} \right)+\ldots & z\to \infty\:\:\:\: (\text{IR}) \\
    \end{cases},
\end{equation}
where $\alpha=d-\Delta$ and $\tilde{\alpha}= \tilde{\Delta}-d$ are both positive. The constants $\mu$ and $\tilde{\mu}$ are mass scales fixed by potential's couplings, $\phi_{UV}$ and $\phi_{IR}$ respectively. These mass scales define the $\text{UV}$ and $\text{IR}$ regimes when $\mu z \ll 1$ and $\tilde{\mu}z \gg 1$.

In General Relativity it is common to impose constraints to the matter content in order to discard non physical solutions. This is the case for the singularity theorems\cite{Kontou:2020bta}. Among the standard pointwise energy conditions the Null Energy Condition (NEC) is the weakest one. It states that $T_{M N}\eta^{M}\eta^{N}\geq 0$ for any null vector $\eta^M=(\eta^z,\eta^0,\vec{\eta})$. Using the metric \eqref{DomainWall} any null vector is constrained by
\begin{equation}
0=g_{M N}\eta^{M}\eta^{N} \:\:\:\:\:\:\Longrightarrow\:\:\:\:\:\: (\eta^0)^2=\frac{1}{f(z)}(\eta^z)^2+(\vec{\eta})^2.
\end{equation}
Given the contraction
\begin{equation}
\begin{split}
T_{M N}\eta^{M}\eta^{N} & =\frac{\ell_{UV}^2}{z^2}\left(\frac{1}{f(z)}T^z_z(\eta^z)^2-T^0_0(\eta^0)^2+T^i_i (\vec{\eta})^2 \right)\\
& =\frac{\ell_{UV}^2}{z^2}\left(\frac{1}{f(z)}(T^z_z-T^0_0)(\eta^z)^2+(T^i_i-T^0_0) (\vec{\eta})^2 \right),
\end{split}
\end{equation}
then, $T^z_z-T^0_0\geq 0$. Moreover, using Einstein's equations
\begin{equation}
T^z_z-T^0_0=R^z_z-R^0_0=\frac{z^2}{\ell_{UV}^2}\left(f(z)R_{zz}+R_{00}\right)=(d-1)\frac{z f'(z)}{2\ell_{\text{UV}}^2}\geq 0\:\:\:\: \Longrightarrow \:\:\:\:f'(z)\geq 0.
\end{equation}
Since for arbitrary holographic RG flows $f(0)=1$ and $f(\infty)=(\ell_{\text{UV}}/\ell_{\text{IR}})^2$, the NEC implies
\begin{equation}
    1 \leq \; f(z)\; \leq \left(\frac{\ell_{\text{UV}}}{\ell_{\text{IR}}}\right)^2,\:\:\:\:\:\:\:\:
    0 \leq \; f'(z).
\end{equation}

\section{Euler-Lagrange approach} \label{sec:Lagrange}

In the main part of the work we calculated explicitly the large radius expansion of the holographic EE using the Hamilton-Jacobi equation. Here we will show how the same results can be derived by solving the equation of motion (\ref{eq:ELeq}) that results from extremizing the area, and evaluating it on-shell. We reproduce the equation here for convenience,
\begin{equation}\label{eq:eom_rho2}
    \rho(z) z^{d-1} \sqrt{\rho'(z)^2+\frac{1}{f(z)}}\partial_z\left(\frac{\rho'(z)}{z^{d-1}\sqrt{\rho'(z)^2+\frac{1}{f(z)}}}\right) - \frac{(d-2)}{f(z)}=0.
\end{equation} 

Th Hamilton Jacobi approach is more direct; on the other hand, solving the Euler Lagrange equations gives a better understanding of the geometry of the minimal surface. So both methods are complementary. We start with the simpler cases of AdS and $d=2$. Then we pass to the general case and study the large $R$ expansion.

\subsection{Simpler cases}

\subsubsection{Pure AdS$_{d+1}$}
This is the case where $f(z)$ is constant for all $z$ and there is no RG flow. If we start with the UV theory, then $f(z)=1$ and the solution to the equation \eqref{eq:ELeq} is  simply
\begin{equation}
\label{eq:rho_UV}
    \rho_{\text{UV}}(z) = \sqrt{z_t^2 - z^2}.
\end{equation}
It satisfies the boundary conditions if 
\be\label{eq:ztUV}
z_t^{UV}=\sqrt{R^2+\epsilon^2}\,.
\ee
The minimal surface is a hemisphere centered at $z=\rho=0$, with radius $z_t^{UV}$. This is the radius at $z=0$, which is different than the size $R$ of the entangling region at $z=\epsilon$.

Similarly, in the IR theory, $f(z)=\ell_{UV}^2/\ell_{IR}^2$ and the solution reads
\begin{equation}
\label{eq:rho_IR}
    \rho_{\text{IR}}(z) = \frac{\ell_{\text{IR}}}{\ell_{\text{UV}}} \sqrt{z_t^2 - z^2}.
\end{equation}
In this case the boundary conditions imply that 
\be\label{eq:ztIR}
z_t^{IR} = \sqrt{\frac{\ell_{UV}^2}{\ell_{IR}^2}\;R^2+\epsilon^2}\,.
\ee
These solutions are valid for all $d$.

Evaluating the UV solution (\ref{eq:rho_UV}) in the holographic EE formula (\ref{eq:Sdef}) we reobtain the solution \eqref{eq:S_UV}
\begin{equation}
    S_{\text{UV}} = \gamma_d  \int_{\epsilon/z_t^{\text{UV}}}^{1}  d\omega\:\:\:\omega^{1-d}(1-\omega^2)^{\frac{d-3}{2}}.
\end{equation}

\subsubsection{Two dimensional case and c-theorem}

When $d=2$ the $\rho(z)$ dependence disappears in the equation (\ref{eq:eom_rho2}) leaving an equation involving just $\rho'(z)$ and $\rho''(z)$  . Therefore the variational problem contains a conserved quantity and the solution can be written as
\begin{equation}
    \rho(z) = R -\int_\epsilon^{z}  \frac{dz}{f(z)^{1/2}} \frac{1}{\sqrt{(z_t/z)^2-1}}.
\end{equation}
The turning point is fixed by
\begin{equation}\label{z_t}
    \rho(z_t)=0 \;\;\Longrightarrow \;\; R = \int_\epsilon^{z_t}  \frac{dz}{f(z)^{1/2}} \frac{1}{\sqrt{(z_t/z)^2-1}}.
\end{equation}
As a consequence of the NEC (\ref{eq:NEC_conseq}), $f(z)\geq 1$, so for an arbitrary $z_t$
\begin{equation} \label{R<z_t}
    R \leq \int_\epsilon^{z_t} dz\; \frac{1}{\sqrt{(z_t/z)^2-1}} = \sqrt{z_t^2- \epsilon^2}.
\end{equation}
From (\ref{eq:ztUV}), $R=\sqrt{\big(z_t^{\text{UV}}\big)^2-\epsilon^2}$, so (\ref{R<z_t}) implies $z_t^{\text{UV}}\leq z_t$. Similarly, because $f(z)\leq \left(\frac{\ell_{UV}}{\ell_{IR}}\right)^2$, we have $z_t\leq z_t^{\text{IR}}$. The combined inequalities give
\begin{equation}
z_t^{UV}\leq z_t\leq z_t^{IR}\,.
\end{equation}

The first derivative of $S$ with respect to $R$ can be written as an expression in terms $\rho(z)$ evaluated at the cut-off $\epsilon$, see \eqref{eq:dSdR}. In particular for $d=2$,
\begin{equation}\label{dSdR_d2}
    R\der{S}{R}=\gamma_2 \frac{R}{z_t}.
\end{equation}
This defines a running c-function by comparing \eqref{dSdR_d2} between the $\text{UV}$ fixed point and the holographic flow, 
\begin{equation}
    \frac{\Delta c}{3}=\frac{c(R)}{3}-\frac{c^{\text{UV}}(R)}{3}= R\der{S}{R}-R\der{S^{\text{UV}}}{R} =  \gamma_2 R \left(\frac{1}{z_t}-\frac{1}{z_t^{\text{UV}}}\right)\leq 0.
\label{eq:c_theo_eulerlag}
\end{equation}
In the large $R$ limit, $z_t\approx \frac{\ell_{UV}}{\ell_{IR}}R$ and the expression extracts the universal coefficient in $d=2$
\begin{equation}\label{DeltaS_d2}
    \frac{\Delta c}{3}=R \der{\Delta S}{R}\Big|_{\tilde{\mu}R\gg 1} = \frac{\ell_{\text{UV}}}{2G_N^{(3)}} -\frac{\ell_{\text{IR}}}{2G_N^{(3)}}  = \frac{(c_{\text{IR}}-c_{\text{UV}})}{3}\leq 0\,,
\end{equation}
where we identify the holographic central charge $c = \frac{3\ell}{2G_N^{(3)}}$ \cite{Brown:1986nw}. Eqs. (\ref{eq:c_theo_eulerlag}) and (\ref{DeltaS_d2}) give a holographic proof of the irreversibility theorem in $d=2$ \cite{Casini:2006es,Freedman:1999gp, Myers:2010xs}.

\subsection{Larger dimensions}

For dimensions $d>2$ the variational problem associated to (\ref{eq:eom_rho2}) does not have a conserved quantity, and the equation of motion cannot be solved analytically for arbitrary $f(z)$. 
Instead, we will follow a matching procedure similar to what was originally done by \cite{Liu:2012eea,Liu:2013una}. 

\begin{figure}[!ht]
    \centering
    \includegraphics[width=0.5\textwidth]{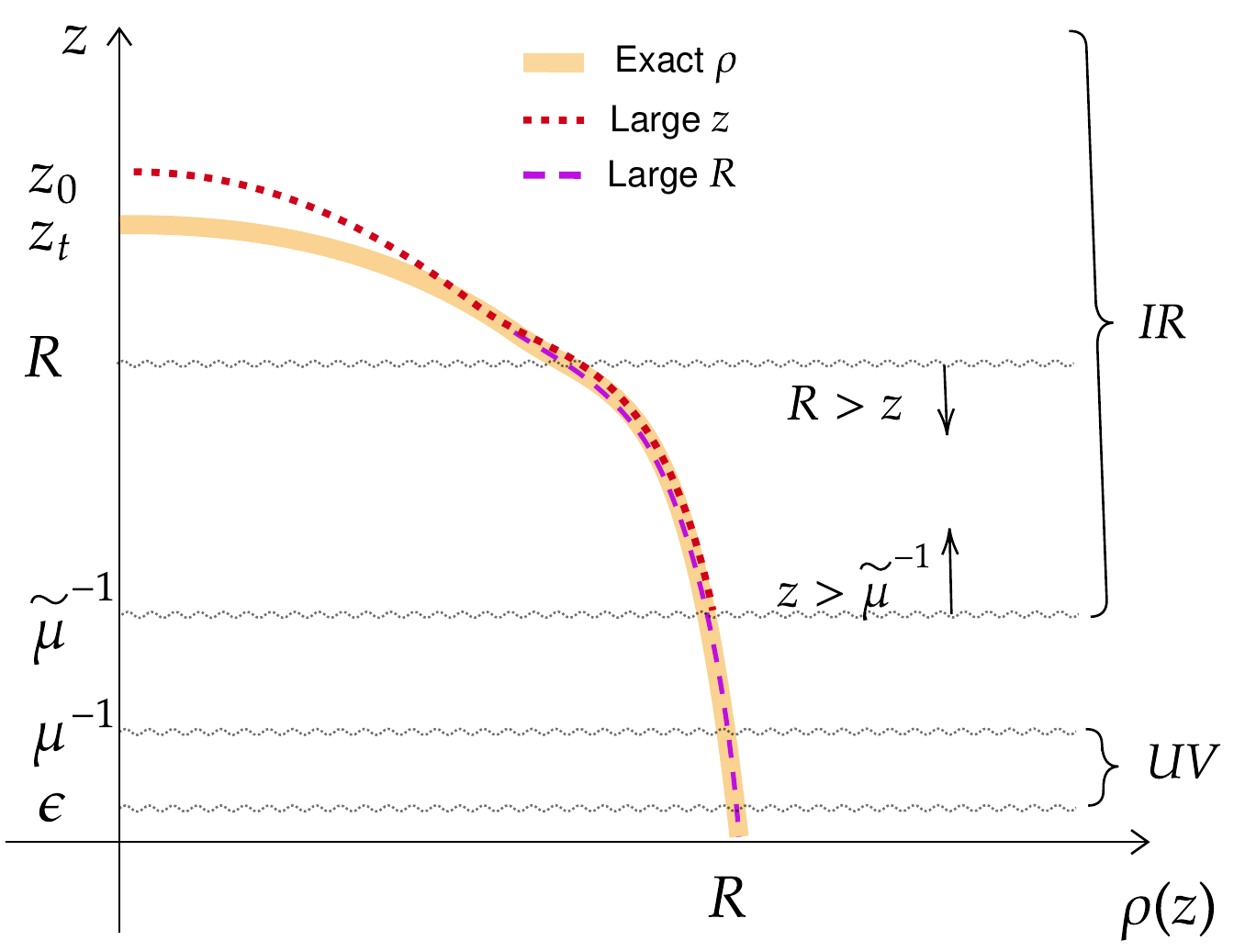}
    \caption{Different approximations for the profile $\rho(z)$ of the minimal surface including all the scales present in the problem. The exact solution of $\rho(z)$,  which is unknown for $d>2$, is depicted in broad plot with turning point $z_t$. In long dashed lines there is a schematic plot of the large $R$ expansion \eqref{IntExpLargeZ} and in short dashed lines there is an schematic plot of the large $z$ expansion (\ref{IRExp}) to order $z_0^{-1}$. The UV is defined for $\epsilon<z<\mu^{-1}$, where $\epsilon$ is an UV cut-off and $\mu$ an UV energy scale; the IR is defined for $\tilde{\mu}<z$ where, $\tilde{\mu}$ is an IR energy scale. The matching procedure is done when $z$ is between $\tilde{\mu}^{-1}<z<R$, where $R$ is the sphere's radius of the spatial region at the boundary.}
    \label{fig:scales}
\end{figure}

The idea is to write two different expansions for $\rho(z)$. The first expansion is written in powers of $R$ in the large $R$ limit, where each coefficient depends on $z$ (Section \ref{sec:largeR}). This expansion is valid near $z=0$ but it is not suitable near the turning point $z_t$, so all coefficients in the expansion have an unknown parameter. The second expansion is written in powers of the $\text{IR}$ mass scale $\tilde{\mu}$. We refer to this as the large $z$ expansion (Section \ref{sec:large_z}). There we can impose the turning point boundary condition but not $\rho(\epsilon)=R$. In the limit where $R$ and $z$ are both large, both expansions overlap and this allows to fix the unknown parameters (Section \ref{sec:matching}). 
In Fig. \ref{fig:scales}, we depict a schematic profile for $\rho(z)$ including all the scales present, and the overlap range.

\subsubsection{Large R expansion}
\label{sec:largeR}

For the large $R$ expansion, let us begin proposing
\begin{equation} \label{firstRho}
    \rho(z,R) = R - \frac{\rho_1(z)}{R^a}.
\end{equation}
This ansatz satisfies the UV boundary condition if $\rho_1(\epsilon)=0$. The minus sign is explicitly chosen because we expect $\rho$ to decrease with $z$ as depicted in Figure \ref{fig:RT}. The linearized equation of motion (\ref{eq:eom_rho2}) for $\rho_1(z)$ is
\begin{equation} \label{firstELeq}
    \frac{z^{d-1}}{\sqrt{f(z)}}\Bigg(\frac{\sqrt{f(z)}}{z^{d-1}} \rho_1'(z)\Bigg)'=-\frac{(d-2)}{f(z)} R^{a-1}.
\end{equation}
To satisfy that $\rho_1(z)$ is independent from $R$ we need $a=1$. The equation can be integrated to obtain the first correction
\begin{equation}\label{rho1}
    \rho_1(z) = b_1 \int_\epsilon^z du \frac{u^{d-1}}{\sqrt{f(u)}} + (d-2) \int_\epsilon^z du \frac{u^{d-1}}{\sqrt{f(u)}} \int_u^\infty dv \frac{1}{v^{d-1}\sqrt{f(v)}}\,.
\end{equation}
The ansatz \eqref{firstRho} is valid for $\frac{\rho_1(z)}{R^2} \ll 1$. Here $b_1$ an arbitrary constant that multiplies the homogeneous solution. This reflects the fact that we cannot impose the IR boundary condition $\rho(z_t)=0$ in this expansion.

The next corrections involve higher odd powers of $1/R$. The general solution also includes non-integer powers of $1/R$ that satisfy the homogeneous equation without source. These terms contribute nonlocal corrections to the EE that go to zero at large $R$, so we will not focus on them. In summary, we have
\begin{equation}\label{largeR}
    \rho(z)= R - \frac{\rho_1(z)}{R} - \frac{\rho_3(z)}{R^3}  - ... - \frac{\rho_\nu(z)}{R^\nu}-\ldots \,.
\end{equation}
Expanding the Euler-Lagrange equation in $1/R$ gives equations for each $\rho_n$,
\begin{equation} \label{eqrhon}
    \frac{z^{d-1}}{\sqrt{f(z)}}\left(\frac{\sqrt{f(z)}}{z^{d-1}} \rho_n'(z)\right)' = s_n(z)\,,
\end{equation}
where $s_n(z)$ is a source term for $\rho_n(z)$, which depends on $\rho_m(z)$ with $m<n$. The first values for the sources are
\begin{equation} \label{sources}
        s_1(z) = - \frac{(d-2)}{f(z)}\;,\;
        s_3(z) = -\rho_1(z) \frac{(d-2)}{f(z)} - (d-2) \rho_1'(z)^2 + \frac{(d-1)}{z} f(z) \rho_1'(z)^3.
\end{equation}
Integrating \eqref{eqrhon} gives the solutions
\begin{equation}\label{rhon}
    \rho_n(z) = b_n \int_\epsilon^z du \frac{u^{d-1}}{\sqrt{f(u)}} - \int_\epsilon^z du \frac{u^{d-1}}{\sqrt{f(u)}} \int_u^\infty dv \frac{\sqrt{f(v)}}{v^{d-1}} s_n(z)\,,
\end{equation}
with the $b_n$ undetermined at this stage.

\subsubsection{Large z expansion}
\label{sec:large_z}

Now we look for approximate solutions of the Euler-Lagrange equation in the IR regime $\tilde{\mu}z \gg 1$. Since $z<z_t$, in this limit we have $\tilde{\mu}z_t\gg 1$. In the IR zone we expect the solution to be similar to the pure $\text{AdS}_{d+1}$ solution. We add a correction $\sigma_1$ that should be suppressed by powers of $\tilde{\mu}z$,
\begin{equation} \label{firstrhoIR}
    \rho (z)= \frac{\ell_{IR}}{\ell_{UV}} \sqrt{z_0^2 - z^2} + \frac{\ell_{IR}}{\ell_{UV}} \sigma_1(z),
\end{equation}
for some constant $z_0$. 

In the strict IR limit $z_0=z_t^{\text{IR}}$ as discussed in (\ref{eq:rho_IR}). Using the IR approximation \eqref{f(z)} of $f(z)$ in (\ref{eq:eom_rho2}), the linearized equation  for the correction becomes
\begin{equation}\label{eqsigma_1}
\begin{split}
    \sqrt{z_0^2-z^2} \Big( (d-2+\tilde{\alpha})z_0^2z-(\tilde{\alpha}-1)z^3 \Big)\frac{1}{(\tilde{\mu}z)^{2\tilde{\alpha}}} =& \: \sigma _1(z) (d-2)z_0^2z + \sigma _1'(z) \Big( 2z^4+(d-3)z_0^2z^2\\
    &\:\:\:\:\:-(d-1)z_0^4 \Big)+ \sigma _1''(z) \Big(z^{5}+z z_0^4-2 z^{3} z_0^2\Big).
   \end{split}
\end{equation}
The solution to this equation is an integral involving hypergeometric functions, but the exact expression is not illuminating. However, we only need the solution at large $R$, which simplifies as we describe next.

The first thing we learn from this equation is that $\sigma_1(z)$ is suppressed by $(\t\mu z)^{-2\t\alpha}$. The value of $z_t$ can be obtained at this order of approximation from the IR condition $\rho(z_t)=0$,
\begin{equation}\label{z_0}
    z_0^2 = z_t^2 + \sigma_1^2(z_t).
\end{equation}
Thus, $z_0\sim z_t$ plus corrections of order  $(\t\mu z_t)^{-2\t\alpha}$. From the pure $\text{AdS}_{d+1}$ case we expect the turning point to be of order $z_t\sim R$; we will check this self-consistently after finding the solution for $\rho(z)$. Since we want to match (\ref{largeR}) with (\ref{firstrhoIR}) we need to expand \eqref{firstrhoIR} in the large $z_0$ limit. The large $z_0$ expansion applied to \eqref{eqsigma_1} gives a linear equation for $\sigma_1(z)$ at order $z_0^{-1}$
\begin{equation}
    \frac{ z}{z_0}  \frac{(d-2+\tilde{\alpha})}{(\tilde{\mu}z)^{2\tilde{\alpha}}} = -(d-1) \sigma _1'(z)+z \sigma _1''(z),
\end{equation}
with solution\footnote{We do not consider the homogeneous solution $\sigma^{\text{hom}}_1(z)$ since the corresponding matching expression is only related with the non integer powers of $R$.}
\begin{equation}\label{sigma_1}
    \sigma_1(z) =  c_0  + \frac{z^2}{2z_0} \frac{1}{(\tilde{\mu}z)^{2\tilde{\alpha}}} \frac{(\tilde{\alpha} +d-2) }{ (\tilde{\alpha} -1) (2 \tilde{\alpha}
   +d-2)} + \mathcal{O}\left(\frac{1}{z_0^2} \right).
\end{equation}

Replacing this solution into \eqref{firstrhoIR} and expanding the square root, we find the double expansion in $z$ and $z^0$ we were looking for \begin{equation}\label{IRExp}
    \rho(z) = \frac{\ell_{IR}}{\ell_{UV}} z_0 - \frac{\ell_{IR}}{\ell_{UV}}\frac{z^2}{2z_0}  \left( 1 - \frac{1}{(\tilde{\mu}z)^{2\t{\alpha}}} \frac{(\tilde{\alpha} +d-2)}{ (\tilde{\alpha} -1) (2 \tilde{\alpha}+d-2)}+\ldots \right)+ c_0 +  \ldots\:\: .
\end{equation}
The first `$\ldots$' refers to additional subleading powers of $(\tilde{\mu}z)^{2\tilde{\alpha}}$ whereas the second `$\ldots$' refers to additional subleading powers of $1/z_0$. Both terms receive extra contributions if we consider higher corrections in \eqref{firstrhoIR}.

\subsubsection{Matching the expansions}
\label{sec:matching}

The $\rho_n$ corrections \eqref{rhon} have $b_n$ as unspecified parameters. To obtain their values we need information from the IR region. For such reason we expand the integrals in \eqref{rhon} for  $\tilde{\mu}z\gg1$ using $f(z)$ \eqref{f(z)} in the IR zone. The idea is to get an expression that looks like (\ref{IRExp}) in order to extract $b_n$.

The first correction \eqref{rho1} has two terms. To expand the first term, we split the integral introducing an arbitrary value $Z$ in the IR zone $\tilde{\mu}Z\gg1$, so that in the interval $(z,Z)$ the IR approximation is valid. The result is independent of $Z$ up to order $(\tilde{\mu}Z)^{-2\tilde{\alpha}}$, so we opt to consider the limit when $Z\to \infty$. Namely,
\begin{equation}
\begin{split}
    \int_\epsilon^z du\; \frac{u^{d-1}}{\sqrt{f(u)}} &\approx \int_\epsilon^Z du\; \frac{u^{d-1}}{\sqrt{f(u)}} +\frac{\ell_{IR}}{\ell_{UV}} \int_Z^z du\; u^{d-1}\left(1+\frac{1}{2}(\tilde{\mu} u)^{-2\tilde{\alpha}}+...\right)\\
    & = \tilde{\Sigma} + \frac{z^d}{d}\frac{\ell_{IR}}{\ell_{UV}} \left(1+\frac{1}{(\t \mu z)^{2\tilde{\alpha}}} \frac{d}{2(d-2\tilde{\alpha})}+...\right),
\end{split}
\end{equation}
where $\tilde{\Sigma}$ is defined as
\begin{equation}
    \tilde{\Sigma} = \lim_{Z\to\infty} \int_\epsilon^Z du\; \frac{u^{d-1}}{\sqrt{f(u)}} - \frac{Z^d}{d}\frac{\ell_{IR}}{\ell_{UV}} \left(1 + \frac{1}{(\t \mu Z)^{2\tilde{\alpha}}} \frac{d}{2(d-2\tilde{\alpha})}+\ldots\right).
\end{equation}
We use the same method for the second term expansion
\begin{equation}
\begin{split}
    -(d-2)\int_\epsilon^z du\; \frac{u^{d-1}}{\sqrt{f(u)}} \int_\infty^u dv\; \frac{1}{v^{d-1}\sqrt{f(v)}} 
     & \approx -(d-2)\int_\epsilon^Z du\; \frac{u^{d-1}}{\sqrt{f(u)}} \int_\infty^u dv\; \frac{1}{v^{d-1}\sqrt{f(v)}} \\ 
     &\: \:\:\:\:\:\:\:+(d-2)\frac{\ell_{IR}^2}{\ell_{UV}^2}\int_Z^zdu\;u^{d-1}\Big(1 + \frac{1}{2(\tilde{\mu}u)^{2\tilde{\alpha}}}\Big)\times\\
     &\:\:\:\:\:\:\:\:\:\:\:\:\:\:\:\:\:\:\:\:\:\times \int_\infty^u dv\; v^{1-d}\Big(1 + \frac{1}{2(\tilde{\mu}v)^{2\tilde{\alpha}}}\Big)\\
     &= \tilde{\Gamma} + \frac{\ell_{IR}^2}{\ell_{UV}^2}\frac{z^2}{2}\Bigg(1 - \frac{1}{(\tilde{\mu}z)^{2\t{\alpha}}} \frac{(\tilde{\alpha} +d-2)}{ (\tilde{\alpha} -1) (2 \tilde{\alpha}+d-2)}+...\Bigg) ,
\end{split}    
\end{equation}
where $\tilde{\Gamma}$ is defined as
\begin{equation}
\begin{split}
    \tilde{\Gamma}&= \lim_{Z\to\infty} \left[-(d-2) \int_\epsilon^Zdu\; \frac{u^{d-1}}{\sqrt{f(u)}}\int_\infty^u \frac{dv}{v^{d-1}\sqrt{f(v)}}
    \right]\\
    &\:\:\:\:\:\:\:\:\:\:\:\:\:\:\:\:\:\:\:\:\:\:\:\:\:\:\:\:\:\:\:\:\:\:\:\:\:\:\:\:\:\:\:\:-\left[ \frac{\ell_{IR}^2}{\ell_{UV}^2}\frac{Z^2}{2}\Bigg(1 - \frac{1}{(\tilde{\mu}Z)^{2\t{\alpha}}} \frac{(\tilde{\alpha} +d-2)}{ (\tilde{\alpha} -1) (2 \tilde{\alpha}+d-2)}+...\Bigg)\right].
\end{split}
\end{equation}
As a result, the following double expansion holds for $\rho(z)$,
\begin{equation}\label{IntExpLargeZ}
\begin{split}
    \rho(z) = R-\frac{1}{R}\Bigg[ b_1 \tilde{\Sigma} + \tilde{\Gamma} +& \frac{\ell_{IR}^2}{\ell_{UV}^2}\frac{z^2}{2}\Big(1 - \frac{1}{(\tilde{\mu}z)^{2\t{\alpha}}} \frac{(\tilde{\alpha} +d-2)}{ (\tilde{\alpha} -1) (2 \tilde{\alpha}+d-2)}+\ldots\Big)\\
    &+ b_1 \frac{\ell_{IR}}{\ell_{UV}} \frac{z^d}{d} \Big( 1 + \frac{1}{(\t{\mu}z)^{2\t{\alpha}}} \frac{d}{2(d-2\t{\alpha})} + \ldots \Big) \Bigg] + \ldots\:\:\:\:.
\end{split}
\end{equation}

Now that we have both expansions \eqref{IntExpLargeZ} and \eqref{IRExp} in the same form we can match the coefficients in them,
\begin{equation}
    R=\frac{\ell_{IR}}{\ell_{UV}}z_0 \;,\;\:\:\:\:\:\:\:\:\:\: \:\:\:\:\:\:b_1=0 \;,\:\:\:\:\:\:\:\:\:\:\:\:\:\:\:\; c_0 = -\frac{\tilde{\Gamma}}{R}.
\end{equation}
It is important to observe in \eqref{IntExpLargeZ} that the correction with the first power of $R$ has two terms that go like $z^2$ and $z^d$. The unknown parameter $b_1$ is the coefficient of the term $\frac{z^d}{R}$. A similar scenario occurs when higher orders are considered in the expansion. In that case $\rho(z)$ looks like
\begin{equation}
\label{eq:rho_double_exp}
    \rho(z)=R- \frac{\ell_{IR}}{\ell_{UV}}\frac{z^d}{d} \left( \frac{b_1}{R} + \frac{b_3}{R^3} + ... \right) - d_2\frac{z^2}{R} - d_4\frac{z^4}{R^3} -...\:\:,
\end{equation}
for some coefficients $d_{2k}$. Then (\ref{eq:rho_double_exp}) suggests that the $b_n$ coefficients must be identified by looking at the terms $\sim z^d/R^n$ in the large $z$ expansion of $\rho(z)$ \eqref{firstrhoIR}. Since $\sigma_1(z)$ is of order $(\tilde{\mu}z)^{-2\t \alpha}$ only the square root term might contribute to $b_n$. Its expansion for large $z$ is
\begin{equation}
    \frac{\ell_{IR}}{\ell_{UV}} \sqrt{z_0^2-z^2} = \frac{\ell_{IR}}{\ell_{UV}} \sum_{k=0}^\infty \binom{1/2}{k} (-1)^k \frac{z^{2k}}{z_0^{2k-1}}.
\end{equation}
Finally,
\begin{equation}
    b_n = \begin{cases}
   \:\:\:\:\:\:\:\:\:\:\: 0 & n \neq d-1 \\
    -d (-1)^{\frac{d}{2}} 
    \binom{1/2}{d/2} \left( \frac{\ell_{IR}}{\ell_{UV}}\right)^{d-1} & n=d-1\:\:\&\:\: d=\text{even}
    \end{cases}.
\end{equation}
To make computations shorter, we work with non-integer dimensions $d$, so we can effectively consider $b_n=0$ for all $n$. Taking the limit to the integer dimensions recovers the correct result.

\subsection{Entropy expansion}
\label{sec:EE_expansion}

Having obtained the approximate solution $\rho(z)$ at large $R$, we are ready to evaluate the entropy.
To make more explicit the $R$ dependence of the entropy using the Euler-Lagrange approach, we rewrite \eqref{eq:Sdef} as
\begin{equation} \label{S-}
    S=\gamma_d R^{d-2} \int_{\epsilon}^{z_t} \frac{dz}{z^{d-1}}\, \left(\frac{\rho(z)}{R}\right)^{d-2}\, \sqrt{\rho'(z)^2 + \frac{1}{f(z)}}.
\end{equation}
Both terms $\frac{\rho(z)}{R}$ and $\rho'(z)$ contain only inverse odd powers of $R$, thus the large $R$ limit of the integrand reproduces the structure \eqref{eq:Sfp}. The upper limit of integration $z_t$ is also $R$-dependent, and this only contributes to finite terms because at first order $z_t \sim R$. So in order to obtain the $\mu_{d-2k}$ it is sufficient to take $z_t \to \infty$.

The first term from this expansion reproduces the area law and coincides with the result of the HJ formalism \eqref{eq:mu_d_2} when $z_t \to \infty$:
\begin{equation}
    S^{(d-2)}= \gamma_d R^{d-2} \int_\epsilon^\infty \frac{dz}{z^{d-1}} \frac{1}{\sqrt{f(z)}}.
\end{equation}

The second term depends on $\rho_1(z)$, so it needs to be evaluated using the solution \eqref{rho1} with $b_1=0$. In particular, we have
\begin{equation}
\rho'_1(z)=(d-2)\frac{z^{d-1}}{\sqrt{f(z)}}\int_z^{\infty}\frac{dv}{v^{d-1}}\frac{1}{\sqrt{f(v)}}.
\end{equation}
This matches the HJ solution \eqref{eq:mu_d_4} when $z_t \to \infty$,
\begin{equation}
\begin{split}
    S^{(d-4)} =&  \gamma_d R^{d-4} \int_\epsilon^\infty \frac{dz}{z^{d-1}} \frac{1}{\sqrt{f(z)}} \left(\frac{1}{2}f(z) \rho_1'(z)^2-(d-2)  \rho_1(z) \right) \\
    =& -\frac{\gamma_d}{2} R^{d-4} \int_\epsilon^\infty \frac{dz}{z^{d-1}} \sqrt{f(z)} \rho_1'(z)^2\\
    =&-\gamma_d \frac{(d-2)^2}{2}R^{d-4}\int_{\epsilon}^{\infty}dz\:z^{d-1}f(z)\left(\int_z^{\infty}\frac{dv}{v^{d-1}}\frac{1}{\sqrt{f(v)}}\right)^2.
\end{split}
\end{equation}
In the second step the equation of motion for $\rho_1(z)$ \eqref{firstELeq} was applied in order to combine both terms,
\begin{equation}
    -\int_\epsilon^\infty dz\;\rho_1(z) \frac{(d-2)}{z^{d-1}\sqrt{f(z)}}  = - \int_\epsilon^\infty \frac{dz}{z^{d-1}} \sqrt{f(z)} \rho_1'(z)^2 + \underbrace{\rho_1(z) \frac{\sqrt{f(z)}\rho_1'(z)}{z^{d-1}}\Bigg|_\epsilon^\infty}_{=0}.
\end{equation}

The procedure can be applied to calculate the higher order terms $\mu_{d-2k}$. In the main text we have instead focused on the HJ method since we found it was more economic than using the minimal surface equation of motion.


\bibliography{EE}{}
\bibliographystyle{utphys}

\end{document}